\documentclass[preprint,aps]{revtex4}

\usepackage{graphicx}

\newcommand{\myfigure}[3]{
	\begin{figure}
	\centerline{
	\includegraphics{#1.ps}}
	\caption{#2}
	\label{#3}
	\end{figure}
}

\begin{document}
\title{Radiating Gravitational Collapse with Shearing Motion and Bulk Viscosity Revisited}
\author{G. Pinheiro and R. Chan}
\affiliation{Observat\'orio Nacional, Coordena\c c\~ao de Astronomia e Astrof\'{\i}sica, \\ 
Rua General Jos\'e Cristino 77, S\~ao Crist\'ov\~ao, CEP 20921--400, \\
Rio de Janeiro, Brazil \\
E-mail: gpinheiro@on.br, chan@on.br}

\begin{abstract}
A new model is proposed to a collapsing star consisting of an
anisotropic fluid with bulk viscosity, radial heat flow and outgoing
radiation.  In a previous paper one of us has introduced a function time dependent
into the $g_{rr}$, besides the time dependent metric functions
$g_{\theta\theta}$ and $g_{\phi\phi}$.  The aim of this work is to generalize
this previous model by introducing bulk viscosity and compare it to the
non-viscous collapse.
The behavior of the density, pressure, mass, luminosity and the
effective adiabatic index is analyzed. Our work is also compared to the case 
of a collapsing fluid with bulk viscosity of another previous model, for a star with
6 $M_{\odot}$.
The pressure of the star, at the beginning of the collapse, is isotropic but
due to the presence of the bulk viscosity the pressure becomes more and more
anisotropic.  The black hole is never formed because the apparent horizon
formation condition is never satisfied, in contrast of the previous model where
a black hole is formed.  An observer at infinity sees a
radial point source radiating exponentially until reaches the time of maximum
luminosity and suddenly the star turns off. In contrast of the former model
where the luminosity also increases exponentially, reaching a maximum and after
it decreases until the formation of the black hole.  The effective adiabatic index 
diminishes due to the bulk viscosity, thus increasing the instability of
the system, in both models, in the former paper and in this work.
\end{abstract}

\date{\today}

\maketitle

\section{Introduction}
One of the most outstanding problems in gravitation theory is the evolution of
a collapsing massive star, after it has exhausted its nuclear fuel.
The problem of constructing physically realistic models for radiating collapsing
stars is one of the aims of the relativistic astrophysics.  However, in order to obtain
realistic models we need to solve complicated systems of nonlinear differential
equations.   In many cases we can simplify the problem considering some restrictions
in these equations and solve the system analytically.  Such models, although
simplified, are useful to construct simple exact models, which are at least
not physically unreasonable.  This allows a clearer analysis of the main physical
effects at play, and it can be very useful for checking of numerical procedures.

A great number of the previous works in gravitational collapse have considered
only shear-free
motion of the fluid \cite{deOliveira85, Bonnor89, Chan89, Chan93}.  
This simplification allows us to
obtain exact solutions of the Einstein's equations in some cases but it is
somewhat
unrealistic.  It is also unrealistic to consider heat flow without viscosity
but
if viscosity is introduced, it is desirable to allow shear in the fluid motion.

In the work \cite{Martinez94} the authors have studied the collapse of a
radiating star with bulk viscosity but they still maintained the shear-free
motion of the fluid.
Thus, it is interesting to study solutions that contains shear,
because it plays a very important role in the study
of gravitational collapse, as shown in \cite{Chan97, Chan98a, Chan00, 
Chan01, Nogueira04, Pinheiro08} and in \cite{Joshi02}.

In the first paper \cite{Chan97,Chan98a} we have compared two collapsing
model: a shear-free and a shearing model.
In this model we have imposed that the metric components $g_{tt}$ and $g_{rr}$
were independent
of the time and only $g_{\theta\theta}$ and $g_{\phi\phi}$ were time dependent.
We were interested in studying the
effect of the shearing motion in the evolution of the collapse.
It was shown
that the pressure of the star, at the beginning of the collapse, is isotropic
but due to the presence of the shear the pressure becomes more and more
anisotropic.
The anisotropy in self-gravitating systems has been reviewed and discussed the
causes for its appearance by Herrera and Santos \cite{Herrera97}.  As shown by
Chan \cite{Chan97, Chan98a} the simplest cause of the presence of anisotropy in a
self-gravitating
body is the shearing motion of the fluid, because it appears without an
imposition ad-hoc \cite{Chan93}.

In the second work \cite{Chan00} we have used the same model of Chan
\cite{Chan97, Chan98a} and we have analyzed a collapsing
radiating star consisting of an anisotropic fluid with shear viscosity undergoing
radial heat flow with outgoing radiation, but without bulk viscosity.

In the third paper \cite{Chan01} we have also used the same model previous papers 
\cite{Chan97} \cite{Chan98a} and we have analyzed a collapsing
radiating star consisting of an anisotropic fluid with bulk viscosity undergoing
radial heat flow with outgoing radiation, but without shear viscosity.

In the fourth work \cite{Chan03} we have generalized our previous models
by introducing a function time dependent into the $g_{rr}$ and in 
a recent paper \cite{Pinheiro08} we have introduced the shear viscosity.

The aim of this work is to generalize our previous model \cite{Chan01}
by introducing a time dependent function into the $g_{rr}$, besides 
the time dependent metric functions $g_{\theta\theta}$ and $g_{\phi\phi}$,
and to compare the physical results with the previous ones. 
 
This work is organized as follows.  In Section 2 we present the Einstein's 
field
equations.  In Section 3 we rederive the junction conditions, since 
in the former paper \cite{Pinheiro08}
have obtained only results without bulk viscosity.  
In Section 4
we present the proposed solution of the field equations.  In Section 5
we describe the model considered in this work for the initial configuration.
In Section 6 we present the energy conditions for a bulk viscous anisotropic fluid.
In Section 7 we show the time evolution of the total mass,
luminosity and the effective adiabatic index and
in Section 8 we summarize the main results obtained in this work.

\section{Field Equations}

We assume a spherically symmetric distribution of fluid undergoing dissipation
in the form of heat flow.  While the dissipative fluid collapses it produces
radiation.  The interior spacetime is described by the most
general spherically symmetric metric, using comoving coordinates,

\begin{equation}
ds^2_{-}  =  -A^2(r,t)dt^2+B^2(r,t)dr^2 
             + C^2(r,t)(d\theta^2+\sin^2 \theta d\phi^2).
\label{eq:dsi}
\end{equation}

The exterior spacetime is described by Vaidya's \cite{Vaidya53} metric, which represents
an outgoing radial flux of radiation,

\begin{equation}
ds^2_{+}=-\left[ 1- {{2m(v)} \over {\bf r}} \right]dv^2-2dvd{\bf r}+{\bf r}^2
(d\theta^2+\sin^2 \theta d\phi^2),
\label{eq:dso}
\end{equation}
where $m(v)$ represents the mass of the system inside the boundary surface
$\Sigma$, function of the retarded time $v$.

We assume the interior energy-momentum tensor is given by

\begin{eqnarray}
G_{\alpha \beta}&=&\kappa T_{\alpha \beta}=\kappa \left[ 
(\mu+p_t)u_{\alpha}u_{\beta}+p_tg_{\alpha \beta}+ 
(p-p_t)X_{\alpha}X_{\beta} \right. \nonumber \\
& & \left. +q_{\alpha}u_{\beta}+q_{\beta}u_{\alpha} - 
 \zeta \Theta \left( g_{\alpha \beta} + u_{\alpha} u_{\beta} \right) \right],
\label{eq:tab}
\end{eqnarray}
where $\mu$ is the energy density of the fluid,
$p$ is the radial pressure, 
$p_t$ is the tangential pressure,
$q^{\alpha}$ is the radial heat flux,
$X_{\alpha}$ is an unit four-vector along the radial direction, 
$u^{\alpha}$ is the four-velocity, 
which have to satisfy $u^{\alpha}q_{\alpha}=0$, 
$X_{\alpha}X^{\alpha}=1$, $X_{\alpha}u^{\alpha}=0$ and $\kappa=8\pi$ 
(i.e., $c=G=1$).
The quantity $\zeta>0$ is the coefficient of bulk viscosity and the
shearing tensor $\sigma_{\alpha \beta}$ is defined as
\begin{equation}
\sigma_{\alpha \beta}=u_{(\alpha;\beta)}+ \dot u_{(\alpha} u_{\beta)}-{1 \over 3} \Theta (g_{\alpha \beta}+u_{\alpha}u_{\beta}),
\label{eq:sab}
\end{equation}
with
\begin{equation}
\dot u_{\alpha}=u_{\alpha;\beta}u^{\beta},
\label{eq:aab}
\end{equation}
\begin{equation}
\Theta=u^{\alpha}_{;\alpha},
\label{eq:theta}
\end{equation}
where the semicolon denotes a covariant derivative and the parentheses in the
indices mean symmetrizations.  

Since we utilize comoving coordinates we have,

\begin{equation}
u^{\alpha}=A^{-1}\delta^{\alpha}_0,
\label{eq:u}
\end{equation}
and since the heat flux is radial
\begin{equation}
q^{\alpha}=q\delta^{\alpha}_1.
\label{eq:qa}
\end{equation}

Thus the non-zero components of the shearing tensor are given by
\begin{equation}
\sigma_{11}={2 B^2 \over {3A}} 
\left( {\dot B \over B} - {\dot C \over C} \right),
\label{eq:sigma11}
\end{equation}
\begin{equation}
\sigma_{22}=-{C^2 \over {3A}}
\left( {\dot B \over B} - {\dot C \over C} \right),
\label{eq:sigma22}
\end{equation}
\begin{equation}
\sigma_{33}=\sigma_{22} \sin^2 \theta.
\label{eq:sigma33}
\end{equation}

A simple calculation shows that
\begin{equation}
\sigma_{\alpha \beta} \sigma^{\alpha \beta}={2 \over {3A^2}} 
 \left( {\dot B \over B} - {\dot C \over C} \right)^2.
\label{eq:sigma2}
\end{equation}

Thus, we define the scalar $\sigma$ as
\begin{equation}
\sigma=-{1 \over {3A}} 
 \left( {\dot B \over B} - {\dot C \over C} \right).
\label{eq:sigma}
\end{equation}

Using (\ref{eq:dsi}) and (\ref{eq:theta}), we can write that
\begin{equation}
\Theta={1 \over A}\left({{\dot B \over B} + 2 {\dot C \over C}}\right).
\label{eq:theta1}
\end{equation}

The non-vanishing components of the field equations, using (\ref{eq:dsi}),
(\ref{eq:tab}), (\ref{eq:u}), (\ref{eq:qa}) and (\ref{eq:theta1}) 
, interior of the boundary
surface $\Sigma$ are

\begin{eqnarray}
G^{-}_{00} &=& -{ \left( A \over B \right) }^2 \left[ 2 {C'' \over C} +
{ \left( C' \over C \right) }^2 - 2{C' \over C}{B' \over B} \right] \nonumber \\
& & + {\left( A \over C \right)}^2 + {\dot C \over C}{ \left( {\dot C \over C}
+ 2{\dot B \over B} \right) } = \kappa A^2\mu,
\label{eq:g00}
\end{eqnarray}

\begin{eqnarray}
G^{-}_{11} &=& {C' \over C} { \left( {C' \over C} + 2{A' \over A} \right) }-
{ \left( B \over C \right) }^2 \nonumber \\
& &- { \left( B \over A \right) }^2
\left[ 2{\ddot C \over C} + { \left( \dot C \over C \right) }^2 -
2{\dot A \over A} {\dot C \over C} \right] \nonumber \\
& &= \kappa B^2 (p - \zeta\Theta),
\label{eq:g11}
\end{eqnarray}

\begin{eqnarray}
G^{-}_{22} &=& 
{ \left( {C \over B} \right) }^2 { \left[ {C'' \over C} + {A'' \over A}+
{C' \over C}{A' \over A} - {A' \over A}{B' \over B} - {B' \over B}{C' \over C}
\right] } \nonumber \\
& & + { \left( C \over A \right) }^2 { \left[ -{\ddot B \over B} -
{\ddot C \over C} - {\dot C \over C}{\dot B \over B} + {\dot A \over A}
{\dot C \over C} + {\dot A \over A}{\dot B \over B} \right] } \nonumber \\
& &= \kappa C^2 (p_t - \zeta \Theta),
\label{eq:g22}
\end{eqnarray}

\begin{eqnarray}
G^{-}_{33} &=& {G^{-}_{22} \sin^2 \theta},
\label{eq:g33}
\end{eqnarray}

\begin{eqnarray}
G^{-}_{01} &=& -2{\dot C' \over C} + 2{C' \over C}{\dot B \over B}+
2{A' \over A}{\dot C \over C}= -\kappa A B^2 q.
\label{eq:g01}
\end{eqnarray}

The dot and the prime stand for differentiation with respect to $t$ and $r$,
respectively.

\section{Junction Conditions}

We consider a spherical surface with its motion described by a time-like
three-space $\Sigma$, which divides spacetimes into interior and exterior
manifolds.  For the junction conditions we follow the approach given by
Israel \cite{Israel66a, Israel66b}.  Hence we have to demand

\begin{equation}
(ds^2_{-})_{\Sigma}=(ds^2_{+})_{\Sigma},
\label{eq:dsidso}
\end{equation}

\begin{equation}
K^{-}_{ij}=K^{+}_{ij},
\label{eq:kijikijo}
\end{equation}
where $K^{\pm}_{ij}$ is the extrinsic curvature to $\Sigma$, given by

\begin{equation}
K^{\pm}_{ij}=-n^{\pm}_{\alpha}
{{\partial^2x^{\alpha}} \over {\partial \xi^i \partial \xi^j}}
-n^{\pm}_{\alpha}\Gamma^{\alpha}_{\beta \gamma}
{{\partial x^{\beta}}  \over {\partial \xi^i}}
{{\partial x^{\gamma}} \over {\partial \xi^j}},
\label{eq:kij}
\end{equation}
and where $\Gamma^{\alpha}_{\beta \gamma}$ are the Christoffel symbols, 
$n^{\pm}_{\alpha}$ the unit normal vectors to $\Sigma$, $x^{\alpha}$ are
the coordinates of interior and exterior spacetimes and $\xi^i$ are the 
coordinates that define the surface $\Sigma$.

From the junction condition (\ref{eq:dsidso}) we obtain

\begin{equation}
{dt \over {d \tau}}=A(r_{\Sigma},t)^{-1},
\label{eq:ts}
\end{equation}

\begin{equation}
C(r_{\Sigma},t)={{\bf r}_{\Sigma}(v)},
\label{eq:cs}
\end{equation}

\begin{equation}
\left( dv \over {d \tau} \right)^{-2}_{\Sigma}=\left( 1 - {2m \over {\bf r}} + 
2 {d{\bf r} \over dv} \right)_{\Sigma},
\label{eq:dvdtau}
\end{equation}
where $\tau$ is a time coordinate defined only on $\Sigma$.

The unit normal vectors to $\Sigma$ (for details see \cite{Santos85}) are given by 

\begin{equation}
n^{-}_{\alpha}=B(r_{\Sigma},t)\delta^1_{\alpha},
\label{eq:ni}
\end{equation}

\begin{equation}
n^{+}_{\alpha}=\left( 1 - {2m \over {\bf r}} + 
2 {d{\bf r} \over dv} \right)^{-1/2}_{\Sigma}
\left( -{{d{\bf r}} \over dv}\delta^0_{\alpha} + 
\delta^1_{\alpha} \right)_{\Sigma}.
\label{eq:no}
\end{equation}

The non-vanishing extrinsic curvature are given by

\begin{equation}
K^{-}_{\tau \tau}=-\left[ {\left( {dt \over {d \tau}} \right)}^2 {A'A \over B}
\right]_{\Sigma},
\label{eq:k00i}
\end{equation}

\begin{equation}
K^{-}_{\theta \theta}= \left( {C'C \over B} \right)_{\Sigma},
\label{eq:k22i}
\end{equation}

\begin{equation}
K^{-}_{\phi \phi}=K^{-}_{\theta \theta} \sin^2 \theta,
\label{eq:k33i}
\end{equation}

\begin{equation}
K^{+}_{\tau \tau}=\left[ {d^2v \over {d \tau^2}} 
{\left( dv \over d \tau \right)}^{-1}
-{\left( dv \over d \tau \right)} {m \over {{\bf r}^2}} \right]_{\Sigma},
\label{eq:k00o}
\end{equation}

\begin{equation}
K^{+}_{\theta \theta}= \left[ {\left( dv \over d \tau \right)}
\left( 1 - {2m \over {\bf r}} \right){\bf r}+{d{\bf r} \over d \tau }{\bf r}
\right]_{\Sigma},
\label{eq:k22o}
\end{equation}

\begin{equation}
K^{+}_{\phi \phi}=K^{+}_{\theta \theta} \sin^2 \theta.
\label{eq:k33o}
\end{equation}

From the equations (\ref{eq:k22i}) and (\ref{eq:k22o}) we have

\begin{equation}
\left[ {\left( dv \over d \tau \right)}
\left( 1 - {2m \over {\bf r}} \right){\bf r}+{d{\bf r} \over d \tau }{\bf r}
\right]_{\Sigma}= \left( {C'C \over B} \right)_{\Sigma}.
\label{eq:k22ik22o}
\end{equation}

With the help of equations (\ref{eq:ts}), (\ref{eq:cs}), (\ref{eq:dvdtau}),
we can write (\ref{eq:k22ik22o}) as
 
\begin{equation}
m=\left\{ {C \over 2}\left[ 1 + {\left( \dot C \over A \right)}^2 -
{\left( C' \over B \right)}^2 \right] \right\}_{\Sigma},
\label{eq:ms}
\end{equation}
which is the total energy entrapped inside the surface $\Sigma$ \cite{Cahill70}.

From the equations (\ref{eq:k00i}) and (\ref{eq:k00o}), using (\ref{eq:ts}),
we have

\begin{equation}
\left[ {d^2v \over {d \tau^2}} 
{\left( dv \over d \tau \right)}^{-1}
-{\left( dv \over d \tau \right)} {m \over {{\bf r}^2}} \right]_{\Sigma}
=-\left( {A' \over {AB}} \right)_{\Sigma}.
\label{eq:k00ik00o}
\end{equation}

Substituting equations (\ref{eq:ts}), (\ref{eq:cs}) and ({\ref{eq:ms}) 
into (\ref{eq:k22ik22o}) we can write

\begin{equation}
\left( {dv \over {d \tau}} \right)_{\Sigma}= 
\left( {C' \over B} + {\dot C \over \ A} \right)^{-1}_{\Sigma}.
\label{eq:dvdtau1}
\end{equation}

Differentiating (\ref{eq:dvdtau1}) with respect to $\tau$ and using equations
(\ref{eq:ms}), (\ref{eq:dvdtau1}), we can rewrite (\ref{eq:k00ik00o}) as

\begin{eqnarray}
\lefteqn{ {\left( C \over {2AB} \right)}_{\Sigma} \left\{ 2 {\dot C' \over C} -
2{C' \over C}{\dot B \over B} - 2 {A' \over A}
{\dot C \over C} + \right. } \nonumber \\
& &\left( B \over A \right) \left[ 2{ \ddot C \over C} - 2{\dot C \over C}
{\dot A \over A} + {\left( A \over C \right)}^2 + 
{ \left( \dot C \over C \right)}^2 - \right. \nonumber \\
& &\left. \left. {\left( A \over B \right)}^2
{\left( C' \over C \right)}^2 - {\left( A \over B \right)}^2
\left( 2{A' \over A}{C' \over C} \right) \right] \right\}_{\Sigma}=0.
\label{eq:pqb}
\end{eqnarray}

Comparing (\ref{eq:pqb}) with (\ref{eq:g11}) and (\ref{eq:g01}), we can 
finally write

\begin{equation}
(p-\zeta \Theta)_{\Sigma}=(qB)_{\Sigma}.
\label{eq:pqbs}
\end{equation}
This result is analogous to the one obtained by Chan \cite{Chan03} for a shearing fluid
motion but now we have generalized for an interior fluid 
with bulk viscosity.

The total luminosity for an observer at rest at infinity is

\begin{equation}
L_{\infty}=-\left({dm \over dv}\right)_{\Sigma}=
-\left[ {dm \over dt} {dt \over {d \tau}} 
{\left( dv \over {d \tau} \right) }^{-1} \right]_{\Sigma}.
\label{eq:ls}
\end{equation}

Differentiating (\ref{eq:ms}) with respect to $t$, using (\ref{eq:ts}),
(\ref{eq:dvdtau1}) and (\ref{eq:g11}), we obtain that

\begin{equation}
L_{\infty}={\kappa \over 2}\left[ (p-\zeta \Theta) C^2 { \left( {C' \over B} + 
{\dot C \over A} \right) }^2 \right]_{\Sigma}.
\label{eq:lsp}
\end{equation}

The boundary redshift can be used to determine the time of formation of
the horizon.  The boundary redshift $z_{\Sigma}$ is given by

\begin{equation}
\left( {dv \over {d\tau}} \right)_{\Sigma}=1+z_{\Sigma}.
\label{eq:zs}
\end{equation}

The redshift, for an observer at rest
at infinity diverges at the time of formation of the black hole.
From (\ref{eq:dvdtau1}) we can see that this happens when

\begin{equation}
{ \left( {C' \over B} + {\dot C \over A} \right) }_{\Sigma}=0.
\label{eq:fbh}
\end{equation}

\section{Solution of the Field Equations}

Again as in former paper \cite{Chan03} we propose solutions of the field equations (\ref{eq:g00})-(\ref{eq:g01})
with the form
\begin{equation}
A(r,t)=A_0(r),
\label{eq:art}
\end{equation}

\begin{equation}
B(r,t)=B_0(r)h(t),
\label{eq:brt}
\end{equation}

\begin{equation}
C(r,t)=rB_0(r)f(t).
\label{eq:crt}
\end{equation}

We have chosen this separation of variables in the metric functions, in order
to have the following properties: (a) when $h(t) \rightarrow 1$ and $f(t) 
\rightarrow 1$ the metric functions represent the static solution of the
initial star configuration; (b) when $h(t) = f(t)$ the metric functions
represent the shear-free solution.
We also remark that, following the junction condition equation (\ref{eq:cs}), 
the function
$C(r_{\Sigma},t)$ represents the luminosity radius of the body as seen by an
exterior observer.  On the other hand, with this solution the proper radius
$\int_0^r B(r,t) dr$ evolve with the time.  Such a property was not present
in the previous models 
\cite{Chan97} \cite{Chan98a} \cite{Chan00} \cite{Chan01} \cite{Nogueira04}.

Thus, the expansion scalar (\ref{eq:theta1}) can be written as
\begin{equation}
\Theta={1 \over {A_0}} \left( {{\dot h \over h} + 2{\dot f \over f}} \right).
\label{eq:thetaa}
\end{equation}

Now the equations (\ref{eq:g00})-(\ref{eq:g01}) can be written as

\begin{equation}
\kappa \mu = \kappa {\mu_0 \over h^2}+ {1 \over {A^2_0}}\left( \dot f \over f \right)
\left( {\dot f \over f} + 2{\dot h \over h} \right)+
{1 \over {r^2B^2_0}}\left( {1 \over f^2} - {1 \over h^2} \right),
\label{eq:mu}
\end{equation}

\begin{eqnarray}
\kappa p&=&\kappa{ p_0 \over h^2} - {1 \over {A^2_0}}\left[ 2{\ddot f \over f}+
{\left( \dot f \over f \right)}^2 \right]-
{1 \over {r^2B^2_0}}\left( {1 \over f^2} - {1 \over h^2} \right)
+{\kappa\zeta \over {A_0}}\left( {{2\dot f \over f}+{\dot h \over h}}\right)
\label{eq:p}
\end{eqnarray}

\begin{eqnarray}
\kappa p_t& = &\kappa {p_0 \over h^2} - {1 \over {A_0^2}} \left( { \ddot f \over f}+
{\ddot h \over h} +{\dot h \over h}{\dot f \over f} \right)+
{{\kappa\zeta} \over {A_0}}\left({2\dot f \over f}+{\dot h \over h}\right)
\label{eq:pt}
\end{eqnarray}

\begin{eqnarray}
\kappa q& = &{2 \over {A_0B^2_0h^2}} \left[ \left( {\dot f \over f} \right)
\left( {B'_0 \over B_0} + {1 \over r} - {A'_0 \over A_0} \right) \right.- \nonumber \\
& & \left.- \left( {\dot h \over h} \right)\left( {B'_0 \over B_0} + {1 \over r} \right) \right],
\label{eq:q}
\end{eqnarray}
where

\begin{equation}
\kappa \mu_0 = -{1 \over {B^2_0}}\left[ 2{B''_0 \over B_0} - 
{\left( B'_0 \over B_0 \right)}^2 + {4 \over r}{B'_0 \over B_0} \right],
\label{eq:mu0}
\end{equation}

\begin{equation}
\kappa p_0 = {1 \over {B^2_0}}\left[ {\left( B'_0 \over B_0 \right)}^2 + 
{2 \over r}{B'_0 \over B_0} + 2{A'_0 \over A_0}{B'_0 \over B_0} +
{2 \over r}{A'_0 \over A_0} \right].
\label{eq:p0}
\end{equation}

We can see from equations (\ref{eq:mu})-(\ref{eq:q}) that when the functions
$h(t)=1$ and $f(t)=1$ we obtain the static perfect fluid configuration.

Substituting equations (\ref{eq:p}) and (\ref{eq:q}) into
(\ref{eq:pqbs}), assuming also that $p_0(r_{\Sigma})=0$,
we obtain a second order differential equation in $h(t)$ and $f(t)$,

\begin{equation}
2{\ddot f \over f} + {\left( \dot f \over f \right)}^2
+{1 \over {h}}\left[a\left( {\dot f \over f} \right)
-\bar a \left({\dot h \over h} \right) \right]+
b\left( {1 \over f^2} - {1 \over h^2} \right)=0,
\label{eq:pqb1}
\end{equation}
where

\begin{equation}
a = \left[ 2{\left(A_0 \over B_0 \right)}
\left( {B'_0 \over B_0} + {1 \over r} - {A'_0 \over A_0} \right)
\right]_{\Sigma},
\label{eq:as}
\end{equation}
\begin{equation}
\bar a = \left[ 2{\left(A_0 \over B_0 \right)}
\left( {B'_0 \over B_0} + {1 \over r} \right)
\right]_{\Sigma},
\label{eq:abs}
\end{equation}
and
\begin{equation}
b = \left( A^2_0 \over {r^2B^2_0} \right)_{\Sigma}.
\label{eq:bs}
\end{equation}

In order to obtain the quantities (\ref{eq:mu})-(\ref{eq:q}) we have first
to find an appropriate $h(t)$ function.  Let us first assume that $f(t)=1$.
Thus, we obtain the differential equation for $h(t)$ given by

\begin{equation}
a_0 \dot h - {h}^2 + 1=0,
\label{eq:ht}
\end{equation}
where $a_0=\bar a/b$ and whose solution is given by \cite{Chan98b}
\begin{equation}
h(t)=-\tanh \left( t \over a_0 \right).
\label{eq:ht1}
\end{equation}

Now, using equation (\ref{eq:ht}) we can write equation (\ref{eq:pqb1}) in the
following way

\begin{equation}
2f\ddot f + {\dot f}^2 + a (f/h) \dot f + b(1-f^2)=0.
\label{eq:ft}
\end{equation}

\myfigure{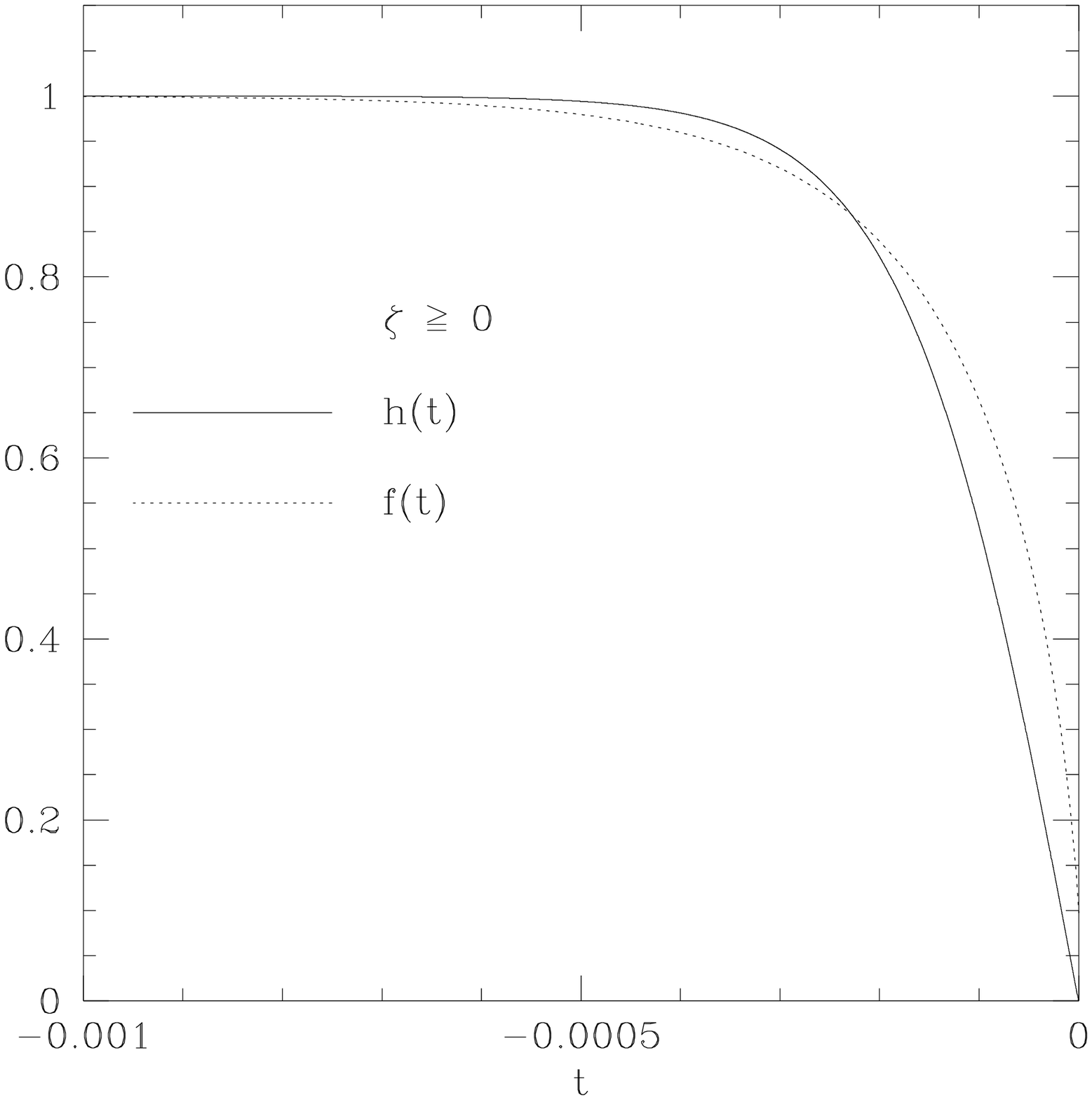}{Time behavior of the functions $f(t)$ and $h(t)$ for the 
model with or without bulk viscosity.  The time is in units of second and 
$f(t)$ and $h(t)$ are dimensionless.  The symbols $\zeta \ge 0$ mean that
the plotted quantity is independent of $\zeta$.}{foft}

\myfigure{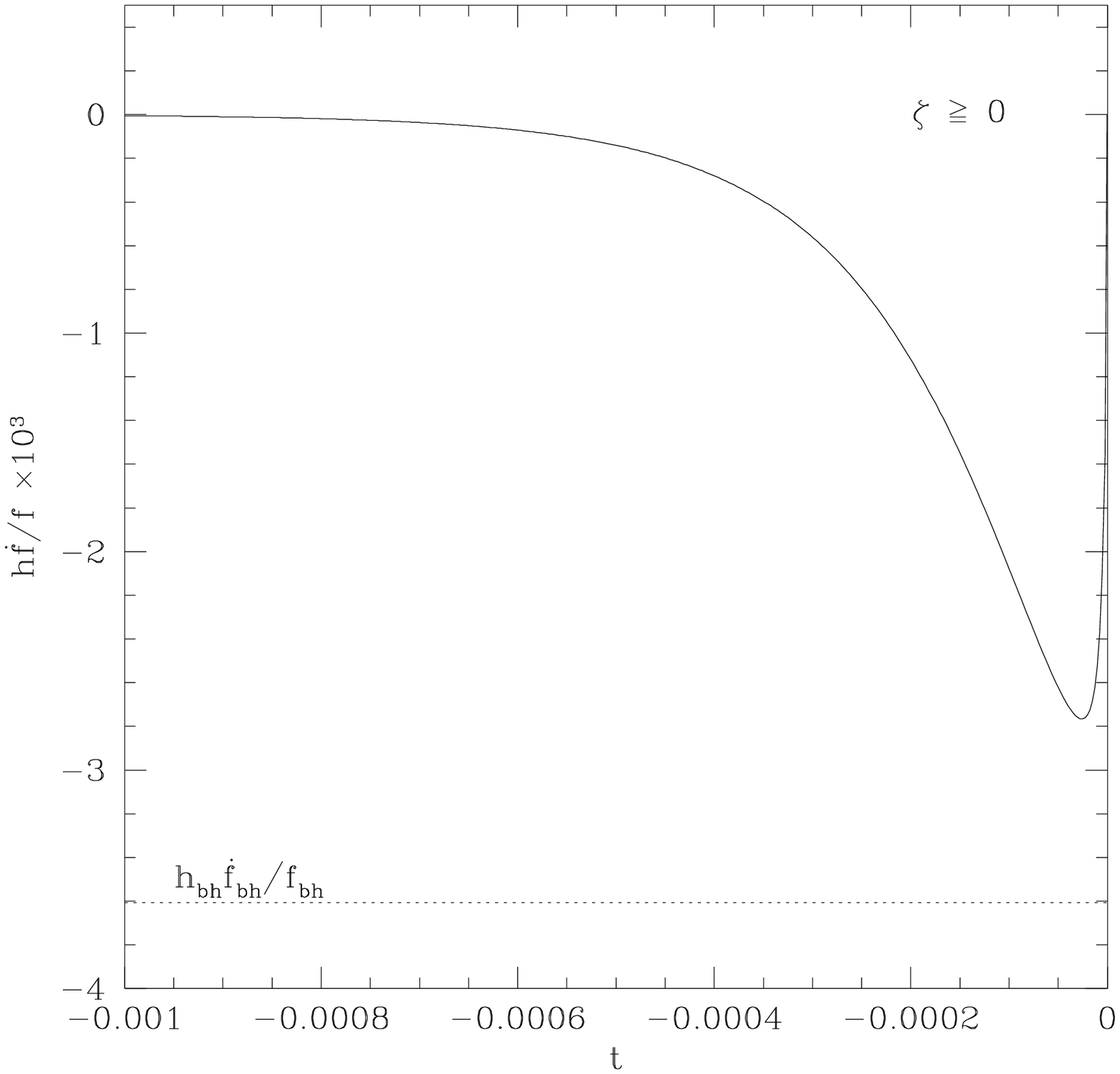}{The function $h\dot f/f$ as a function of the time.
The time is in units of second. The symbols $\zeta \ge 0$ mean that
the plotted quantity is independent of $\zeta$.}{fbh}

\myfigure{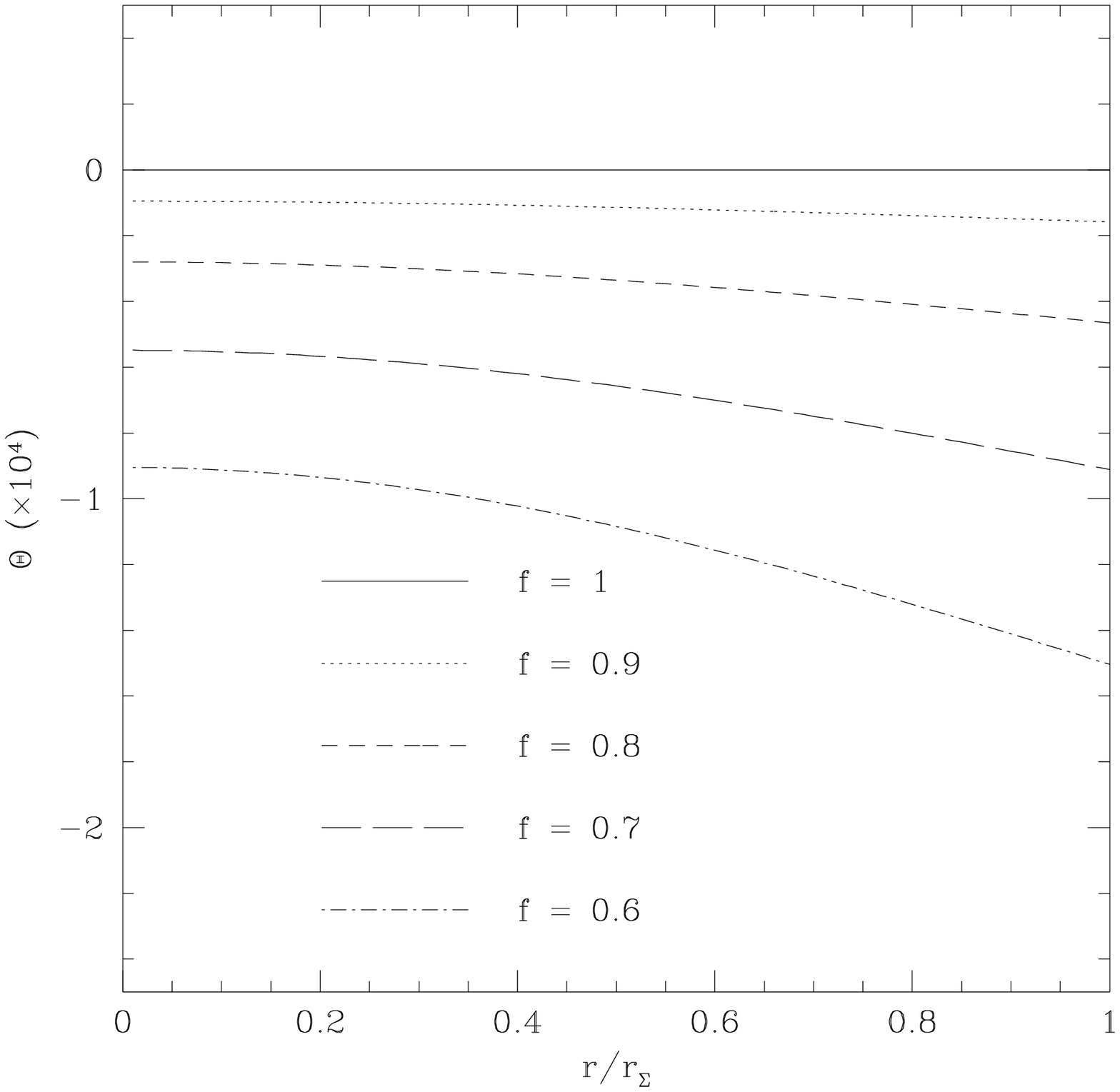}{The expansion scalar profiles as a function of the time.
The radial coordinates $r$ and $r_{\Sigma}$ are in units of second.}{Theta}

\myfigure{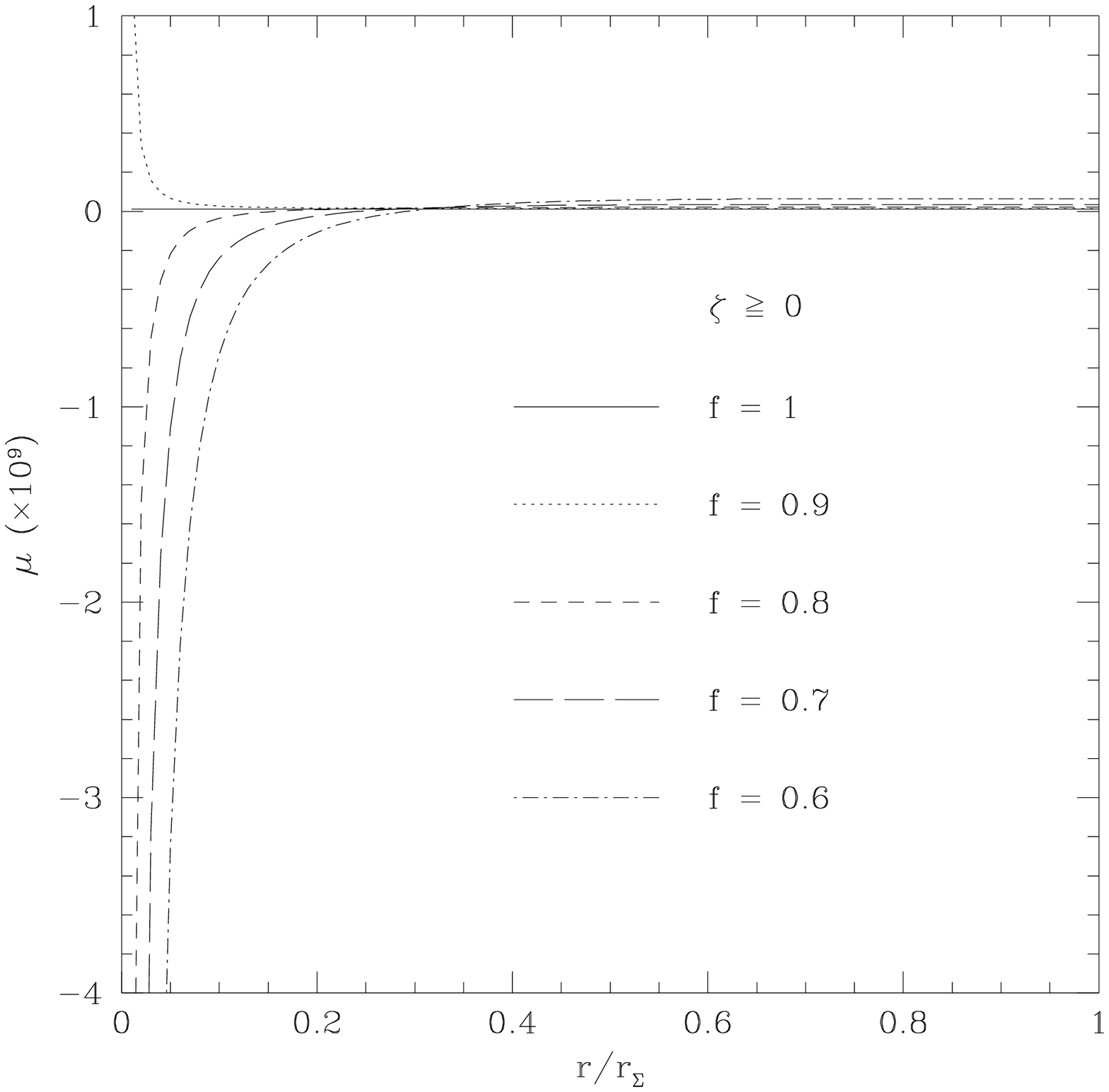}{Density profiles for the model with or without bulk 
viscosity.  The radial coordinates $r$ and $r_{\Sigma}$ are in units of seconds and the 
density is in units of sec$^{-2}$. The symbols $\zeta \ge 0$ mean that
the plotted quantity is independent of $\zeta$.}{mu}

\myfigure{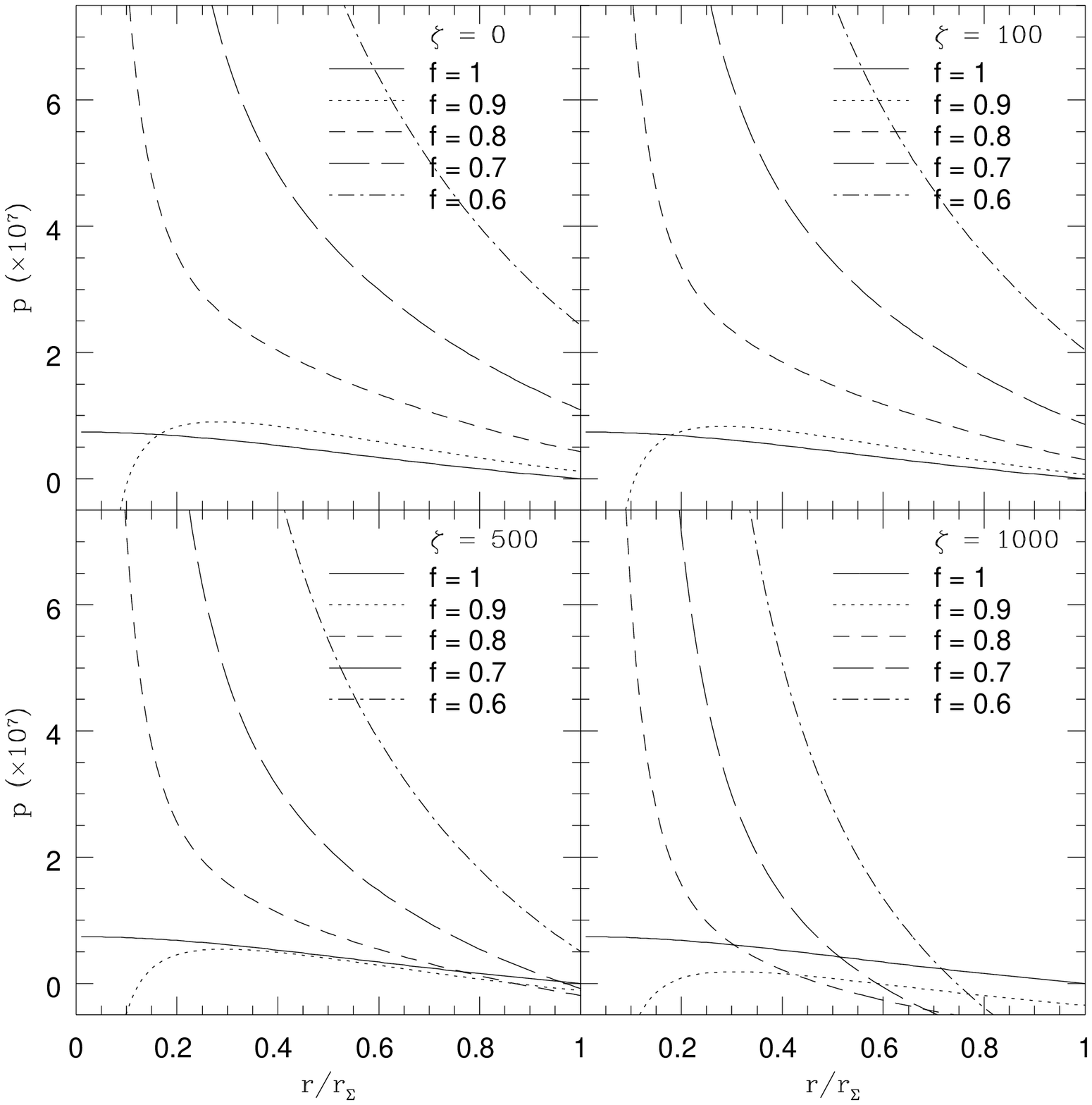}{Radial pressure profiles for four different values of 
$\zeta$ .  The radial coordinates $r$ and $r_{\Sigma}$ are in units of seconds and the 
radial pressure is in units of sec$^{-2}$.}{pr}

\myfigure{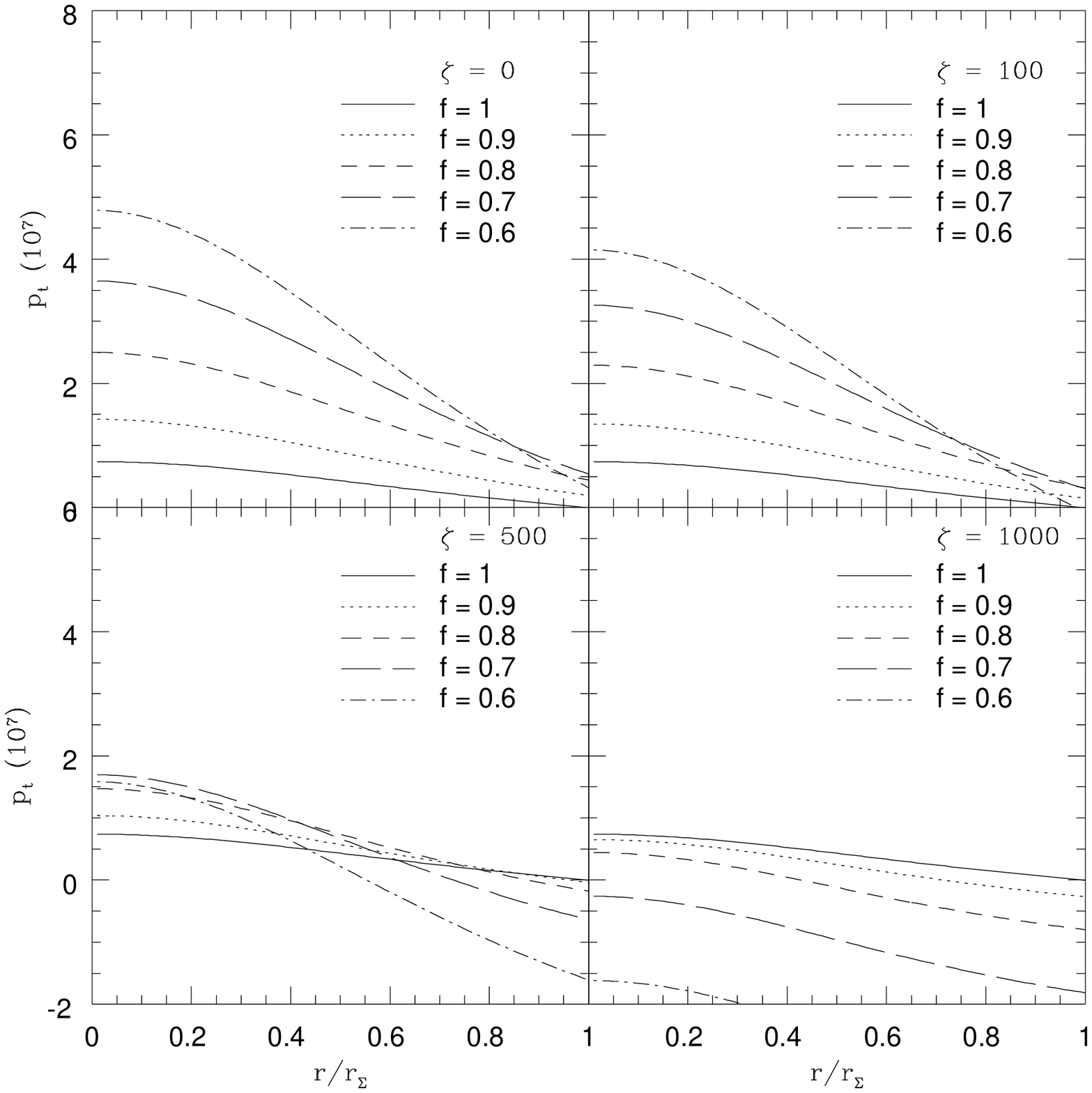}{Tangential pressure profiles for four different values of
$\zeta$ .  The radial coordinates $r$ and $r_{\Sigma}$ are in units of seconds and the 
tangential pressure $p_t$ is in units of sec$^{-2}$.}{pt}

\myfigure{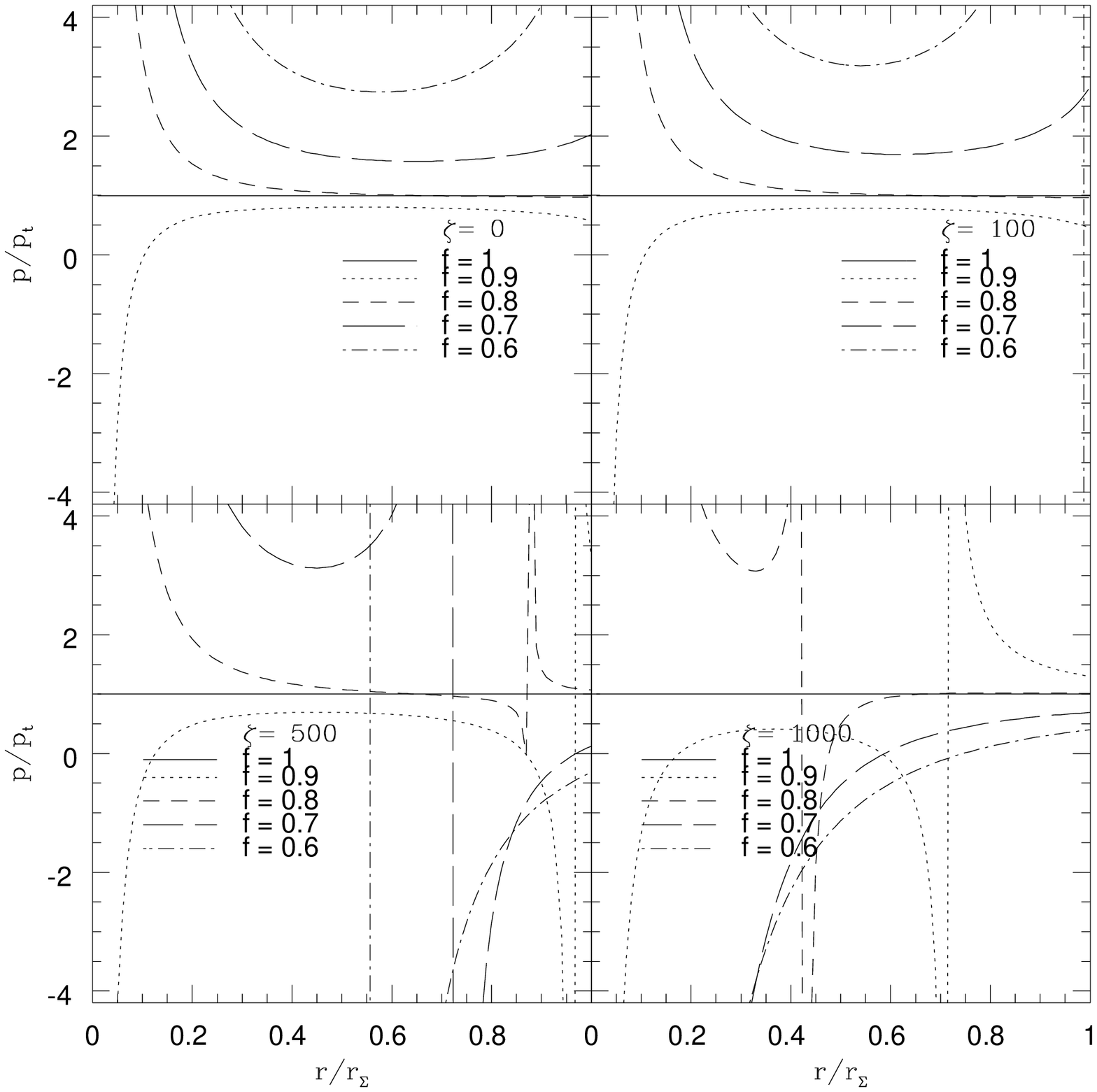}{The profiles for four different values of
$\zeta$ of the ratio between the radial and tangential pressures. The radial coordinates
$r$ and $r_{\Sigma}$ are in units of seconds; and the radial and tangential 
pressure,$p$ and $p_t$, are in units of sec$^{-2}$.}{prpt}

\myfigure{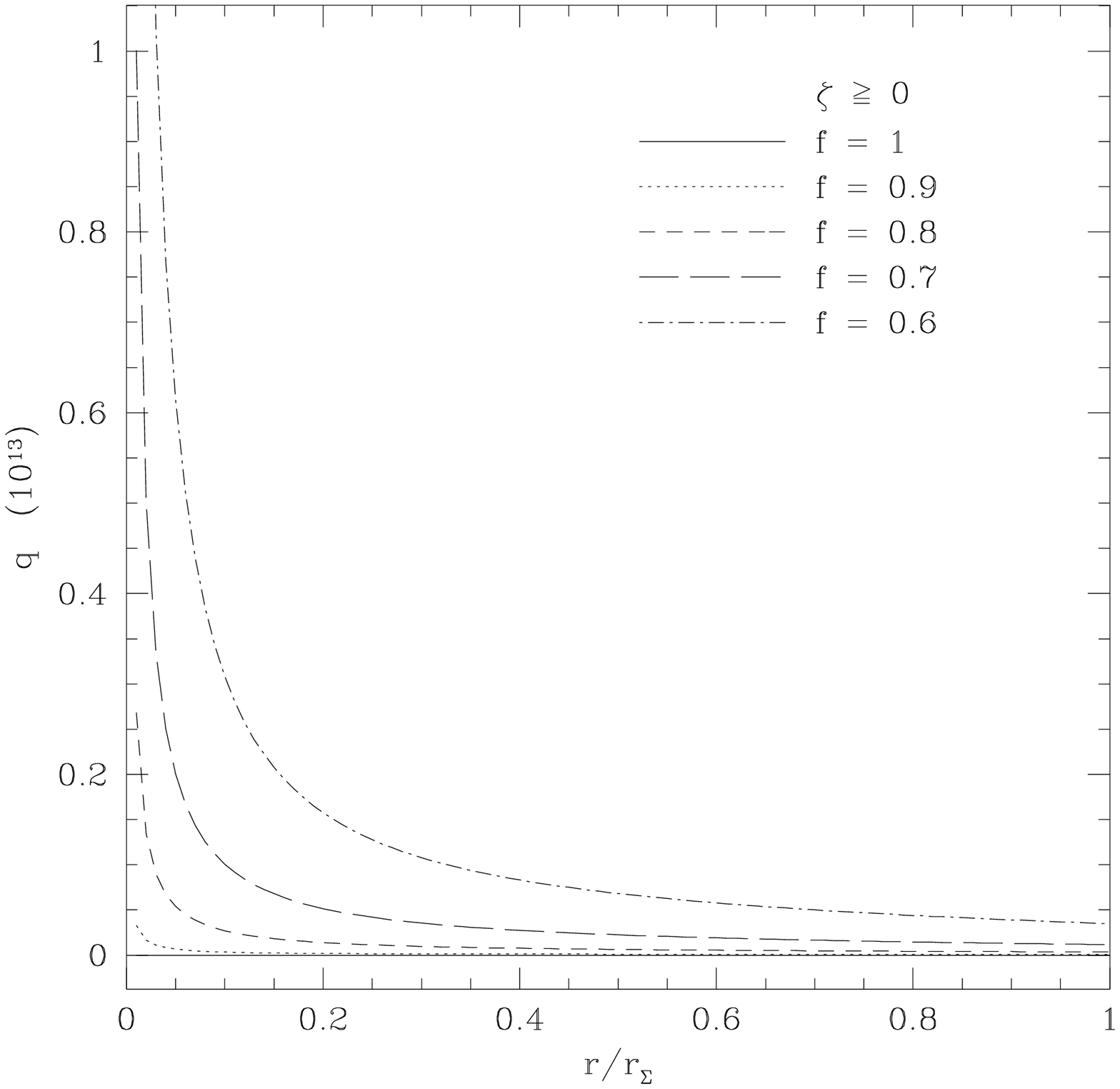}{Heat flux scalar profiles for the model with or without 
bulk  viscosity. The radial coordinate $r$ and $r_{\Sigma}$ are in units of seconds and 
the heat flux $q$ is in units of sec$^{-2}$. The symbols $\zeta \ge 0$ mean that
the plotted quantity is independent of $\zeta$.}{qq}

This equation is almost identical to the one obtained previously 
\cite{Chan97} \cite{Chan98a} \cite{Chan00} \cite{Chan01} \cite{Nogueira04}, 
except the factor $(1/h)$ in the third term.
Thus, as before
it has to be solved numerically, assuming that at
$t\rightarrow-\infty$ represents the static configuration with
$\dot f (t\rightarrow-\infty)\rightarrow0$ and
$f(t\rightarrow-\infty)\rightarrow1$.  We also assume that $f(t\rightarrow 0)
\rightarrow 0$.
This means that the luminosity radius $C(r_{\Sigma},t)$ has the value
$r_{\Sigma}B_0(r_{\Sigma})$ at the beginning of the collapse and vanishing at
the end of the evolution.  Analogously, the proper radius has the value
$h(t)\int_0^r B_0(r) dr$ at the beginning of the collapse and also vanishing at the
end of the collapse.

\section{Model of the Initial Configuration}

We consider that the system at the beginning of the collapse has a static
configuration of a perfect fluid satisfying the Schwarzschild interior
solution \cite{Raychaudhuri79}

\begin{equation}
A_0={g(r) \over {2(1+r^2_{\Sigma})(1 + r^2)}},
\label{eq:a0}
\end{equation}

\begin{equation}
B_0={2R \over {1 + r^2}},
\label{eq:b0}
\end{equation}
where

\begin{equation}
g(r)=3(1-r^2_{\Sigma})(1+r^2)-(1+r^2_{\Sigma})(1-r^2),
\label{eq:gr}
\end{equation}
and

\begin{equation}
R=m_0 {{(1 + r^2_{\Sigma})^3} \over {4r^3_{\Sigma}}}.
\label{eq:rr}
\end{equation}
and where $r_{\Sigma}$ is the radial coordinate relative to 
the physical initial radius of the star in comoving coordinates
and $m_0$ is the initial mass of the system.
Thus the static uniform energy density and static pressure are given by

\begin{equation}
\kappa \mu_0={3 \over R^2},
\label{eq:mu0a}
\end{equation}

\begin{equation}
\kappa p_0={6 \over R^2}{{(r^2_{\Sigma}-r^2)} \over g(r)}.
\label{eq:p0a}
\end{equation}

We consider the initial configuration as due to a iron core of a
presupernova with
$m_0=6M_{\odot}$, initial radial coordinate $r_{\Sigma}=1.6\times10^5$ km, 
which correspond to $2.963\times10^{-5}$ and
$5.337\times10^{-1}$, respectively, in units of second.  Thus, the physical radius
$r_{\Sigma}B_0(r_{\Sigma})=25.742$ km, which gives a density of
$1.675 \times 10^{14}$ g cm$^{-3}$ \cite{Woosley88}. With these
values we can solve numerically the differential equation (\ref{eq:ft}).
We can see from (\ref{eq:q}), using (\ref{eq:a0})-(\ref{eq:rr}) and this
initial configuration, that
$[(B'_0/B_0+1/r-A'_0/A_0)/A_0]_{\Sigma} < 0$, $(B'_0/B_0+1/r)_{\Sigma} > 0$,
$\dot g < 0$
and by the fact that $q_{\Sigma} > 0$ then we conclude that $\dot f < 0$.
In figure \ref{foft} we can see the time evolution of the functions $f(t)$
and $h(t)$.

In order to determine
the time of formation of the horizon $f_{\rm bh}$, we use the equations
(\ref{eq:fbh}), (\ref{eq:art})-(\ref{eq:crt}), (\ref{eq:a0})-(\ref{eq:p0a})
and write

\begin{equation}
{\dot f_{\rm bh} \over f_{\rm bh}} h_{\rm bh} =
- {{2r_{\Sigma}^2(1-r_{\Sigma}^2)^2} \over {m_0(1+r_{\Sigma}^2)^4}}
\approx -3.606\times10^3.
\label{eq:fbh1}
\end{equation}

Using the numerical solution of $f(t)$, $h(t)$ and equation (\ref{eq:fbh1}),
we can see from figure \ref{fbh} that the horizon is never formed, because
the function $h\dot f/f$ does not reach the value $-3.606\times10^3$.
At the first sight this fact could be interpreted as the formation of a
naked singularity.  However, this is not the case as we will see below in
the calculation of the total energy entrapped inside the hypersurface
$\Sigma$.

We will assume that $\zeta$ is constant, but in general the bulk
viscosity coefficient depends on the temperature and density of the
fluid \cite{Cutler87}. The dependence of the expansion scalar on the
time and radial coordinate is shown in the figure \ref{Theta}. 
Hereinafter, the values of $\zeta$ will be $1.347\times 10^{30}$,
$6.736\times 10^{30}$ and $1.347\times 10^{31}$ g cm$^{-1}$ s$^{-1}$,
which correspond to values 100, 500 and 1000 s$^{-1}$, respectively, in time
units.
These values are about ten orders of magnitude above current estimates of
the bulk  viscosity coefficient in neutron stars \cite{Andersson}.  If these
lower values were used in our model, we would have obtained results like the
ones with $\zeta \approx 0$, i.e., without bulk  viscosity.

It is shown in figures \ref{mu} and \ref{qq} the radial profiles of the density
and the heat flux.  It is shown only one plot for each quantity because they
do not depend on the bulk viscosity, which can be seen from equations
(\ref{eq:mu}) and (\ref{eq:q}).

In figure \ref{pr} and \ref{pt} we notice that the radial and tangential pressures 
diminish with the bulk viscosity.

In the figure \ref{prpt} ($\zeta = 0$) we can see that the star is isotropic
at the beginning of the collapse ($f = 1$) but becoming more and more
anisotropic at later times.  

\section{Energy Conditions for a Viscous Anisotropic Fluid}

Following the same procedure used in Kolassis, Santos and Tsoubelis \cite{Kolassis88} we
can generalize the energy conditions for a viscous anisotropic fluid.

For the energy-momentum tensor Segre type $[111,1]$ and if $\lambda_0$ denotes the
eigenvalue corresponding to the timelike eigenvector, the general energy conditions are
equivalent to the following relations between the eigenvalues of the energy-momentum tensor:

\bigskip
\noindent a) weak energy condition
\begin{equation}
-\lambda_0 \ge 0,
\label{eq:wc1}
\end{equation}
and
\begin{equation}
-\lambda_0 +\lambda_i\ge 0,
\label{eq:wc2}
\end{equation}

\bigskip
\noindent b) dominant energy condition
\begin{equation}
\lambda_0 \le \lambda_i \le -\lambda_0,
\label{eq:dc1}
\end{equation}

\bigskip
\noindent c) strong energy condition
\begin{equation}
-\lambda_0 + \sum_i \lambda_i \ge 0,
\label{eq:sc1}
\end{equation}
and
\begin{equation}
-\lambda_0 +\lambda_i\ge 0,
\label{eq:sc2}
\end{equation}
where the values $i=1,2,3$ represent the eigenvalues corresponding to the spacelike
eigenvectors.

The eigenvalues $\lambda$ of the energy-momentum tensor are the roots of the equation
\begin{equation}
\left| T_{\alpha \beta} - \lambda g_{\alpha \beta} \right| = 0.
\label{eq:tablg}
\end{equation}

Thus, we can rewrite equation (\ref{eq:tablg}) as
\[ \begin{array}{c}
\left|
\begin{array}{cccc}
A^2(\mu+\lambda)  &     -AB\bar q                       &        0                     &       0      \\
    -AB\bar q       & B^2(p-\lambda-\zeta\Theta)) &        0                     &       0      \\
     0       &      0                       & C^2(p_t-\lambda-\zeta\Theta) &       0      \\
     0       &      0                       &        0                     & C^2(p_t-\lambda-\zeta\Theta)
\label{eq:det1}
\end{array}
\right| = 0,
\end{array} \]
where $\bar q = q B$ and the determinant of this equation is given by
\begin{eqnarray}
& &\left[ (\mu+\lambda)(\lambda-p+\zeta\Theta)+\bar q^2 \right] \times \nonumber\\ 
& &(\lambda-p_t+\zeta\Theta)^2A^2B^2C^4=0.
\label{eq:det2}
\end{eqnarray}

Thus, one of the solutions of the equation (\ref{eq:det2}) is
\begin{equation}
\left[ (\mu+\lambda)(\lambda-p+\zeta\Theta)+\bar q^2 \right]=0,
\label{eq:det3}
\end{equation}
which can be rewritten as
\begin{equation}
\lambda^2 + (\mu-p+\zeta\Theta)\lambda+\bar q^2-\mu(p-\zeta\Theta)=0.
\label{eq:det4}
\end{equation}

The two roots of the equation (\ref{eq:det4}) are
\begin{equation}
\lambda_0 = - {1 \over 2} (\mu - p +\zeta\Theta+\Delta),
\label{eq:lambda0}
\end{equation}
and
\begin{equation}
\lambda_1 = - {1 \over 2} (\mu - p +\zeta\Theta-\Delta),
\label{eq:lambda1}
\end{equation}
where
\begin{equation}
\Delta^2=(\mu+p-\zeta\Theta)^2-4\bar q^2 \ge 0,
\label{eq:delta2}
\end{equation}
must be greater or equal to zero in order to have real solutions.  This
equation can be rewritten as
\begin{equation}
|\mu+p-\zeta\Theta|-2|\bar q| \ge 0.
\label{eq:delta2a}
\end{equation}

The second solution of the equation (\ref{eq:det2}) is
\begin{equation}
(\lambda-p_t+\zeta\Theta)^2=0,
\label{eq:det5}
\end{equation}
whose roots are given by
\begin{equation}
\lambda_2=\lambda_3=p_t-\zeta\Theta.
\label{eq:lambda2}
\end{equation}

\subsection{Weak Energy Conditions}

From equations (\ref{eq:wc1}) and (\ref{eq:lambda0}) we get the first weak energy
condition written as
\begin{equation}
\mu-p+\zeta\Theta+\Delta \ge 0.
\label{eq:wcond1}
\end{equation}

From equation (\ref{eq:wc2}), setting $i=1$ and using equations (\ref{eq:lambda0})
and (\ref{eq:lambda1}) we get the second weak energy condition given by
\begin{equation}
\Delta \ge 0,
\label{eq:wcond2}
\end{equation}
which is equal to the condition (\ref{eq:delta2}).

From equation (\ref{eq:wc2}), now setting $i=2,3$ (since $\lambda_2=\lambda_3$) and 
using equations (\ref{eq:lambda0})
and (\ref{eq:lambda2}) we get the third weak energy condition given by
\begin{equation}
\mu-p+2p_t-\zeta\Theta+\Delta \ge 0.
\label{eq:wcond3}
\end{equation}

\subsection{Dominant Energy Conditions}

From equation (\ref{eq:dc1}), setting $i=1$ and using equations (\ref{eq:lambda0})
and (\ref{eq:lambda1}) we get the inequality
\begin{equation}
-(\mu-p+\zeta\Theta+\Delta)  \le  -(\mu-p+\zeta\Theta-\Delta) 
 \le  \mu-p+\zeta\Theta+\Delta,
\label{eq:dcond0}
\end{equation}
which can be split into two inequalities, given by
\begin{equation}
\Delta \ge 0,
\label{eq:dcond1}
\end{equation}
and
\begin{equation}
\mu-p+\zeta\Theta \ge 0.
\label{eq:dcond2}
\end{equation}

From equation (\ref{eq:dc1}), setting $i=2,3$ and using equations (\ref{eq:lambda0})
and (\ref{eq:lambda2}) we get the inequality
\begin{equation}
-(\mu-p+\zeta\Theta+\Delta) \le 2(p_t-\zeta\Theta) \le \mu-p+\zeta\Theta+\Delta,
\label{eq:dcond3}
\end{equation}
which again we can split it into two inequalities, given by
\begin{equation}
\mu-p+2p_t-\zeta\Theta+\Delta \ge 0,
\label{eq:dcond4}
\end{equation}
and
\begin{equation}
\mu-p-2p_t+3\zeta\Theta+\Delta \ge 0.
\label{eq:dcond5}
\end{equation}

\subsection{Strong Energy Conditions}

Substituting equations (\ref{eq:lambda0}), (\ref{eq:lambda1}) 
and (\ref{eq:lambda2}) into equation (\ref{eq:sc1}) we get the first strong energy 
condition given by
\begin{equation}
2p_t-2\zeta\Theta+\Delta \ge 0.
\label{eq:scond1}
\end{equation}

Since one of the weak energy conditions, equation (\ref{eq:wc2}), is the same 
for the strong energy condition [equation (\ref{eq:sc2})], thus we have that the second and third 
strong energy conditions are equal to equations (\ref{eq:wcond2})-(\ref{eq:wcond3}),
given by
\begin{equation}
\Delta \ge 0,
\label{eq:scond2}
\end{equation}
and
\begin{equation}
\mu-p+2p_t-\zeta\Theta+\Delta \ge 0.
\label{eq:scond3}
\end{equation}

\subsection{Summary of the Energy Conditions}

Summarizing the results, we rewrite the energy conditions.  The energy
conditions for a spherically symmetric fluid whose energy-momentum tensor is
given by equation (\ref{eq:tab}) are fulfilled if the following inequalities
are satisfied:

\begin{equation}
(i)~~~~~~ |\mu+p-\zeta\Theta|-2|\bar q| \ge 0,
\label{eq:gcr1}
\end{equation}

\begin{equation}
(ii)~~~~ \mu-p+2p_t+\Delta-\zeta\Theta \ge 0,
\label{eq:gcr2}
\end{equation}

\noindent and besides,
\bigskip

\noindent a) for the weak energy conditions

\begin{equation}
(iii)~~~~ \mu-p+\Delta+\zeta\Theta \ge 0,
\label{eq:wcr1}
\end{equation}

\noindent b) for the dominant energy conditions

\begin{equation}
(iv)~~~~~ \mu-p+\zeta\Theta \ge 0,
\label{eq:dcr1}
\end{equation}

\begin{equation}
(v)~~~~ \mu-p-2p_t+\Delta+3\zeta\Theta \ge 0,
\label{eq:dcr2}
\end{equation}

\noindent c) for the strong energy conditions

\begin{equation}
(vi)~~ 2p_t+\Delta-2\zeta\Theta \ge 0,
\label{eq:scr1}
\end{equation}
where $\Delta=\sqrt{(\mu+p-\zeta\Theta)^2-4\bar q^2}$.

\myfigure{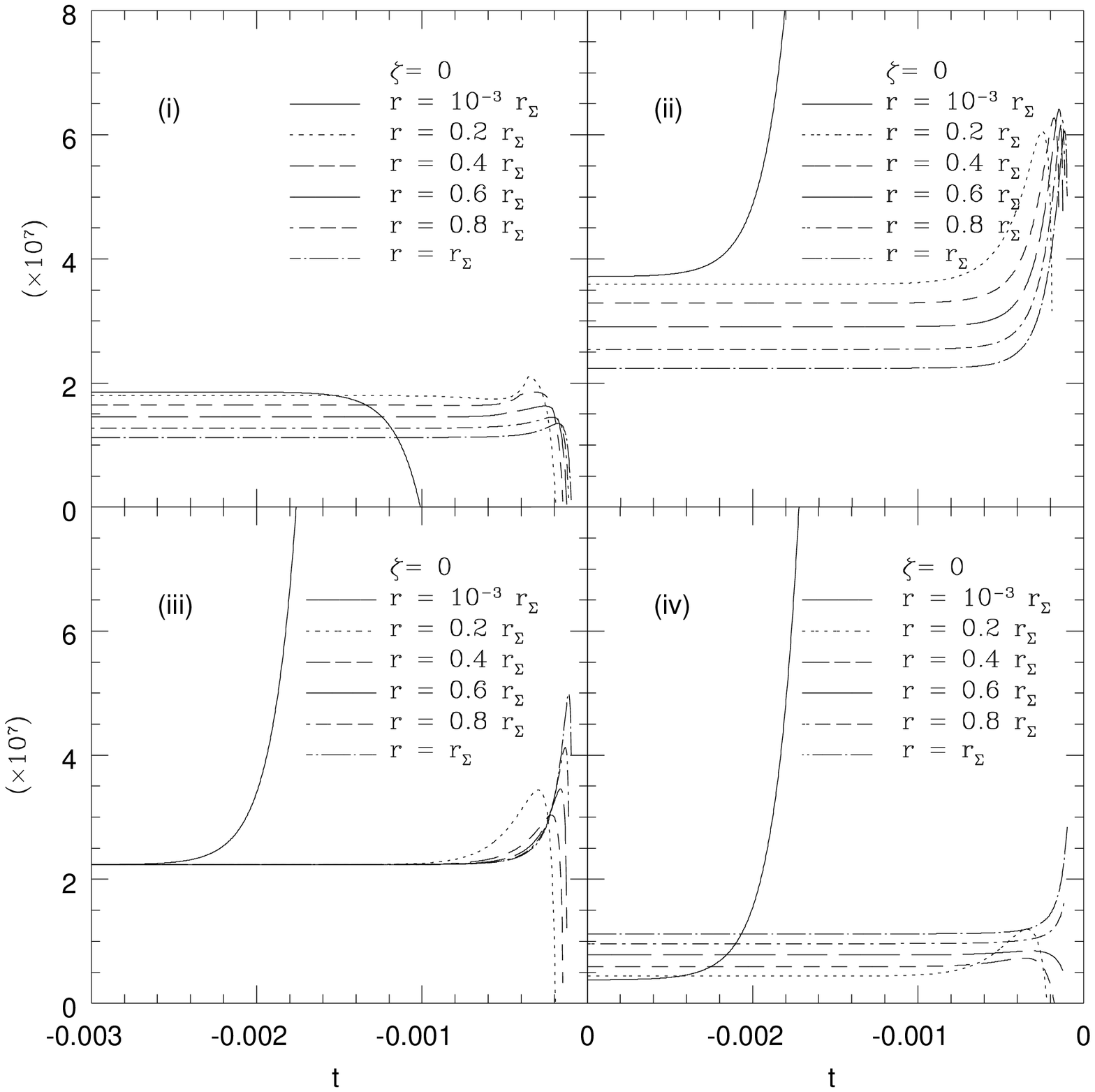}{The energy conditions (\ref{eq:gcr1})-(\ref{eq:dcr1}), 
for the model without bulk viscosity, where $\zeta=0$.  The time is in units 
of seconds and all the others quantities are in units of sec$^{-2}$.}{i-iva}

\myfigure{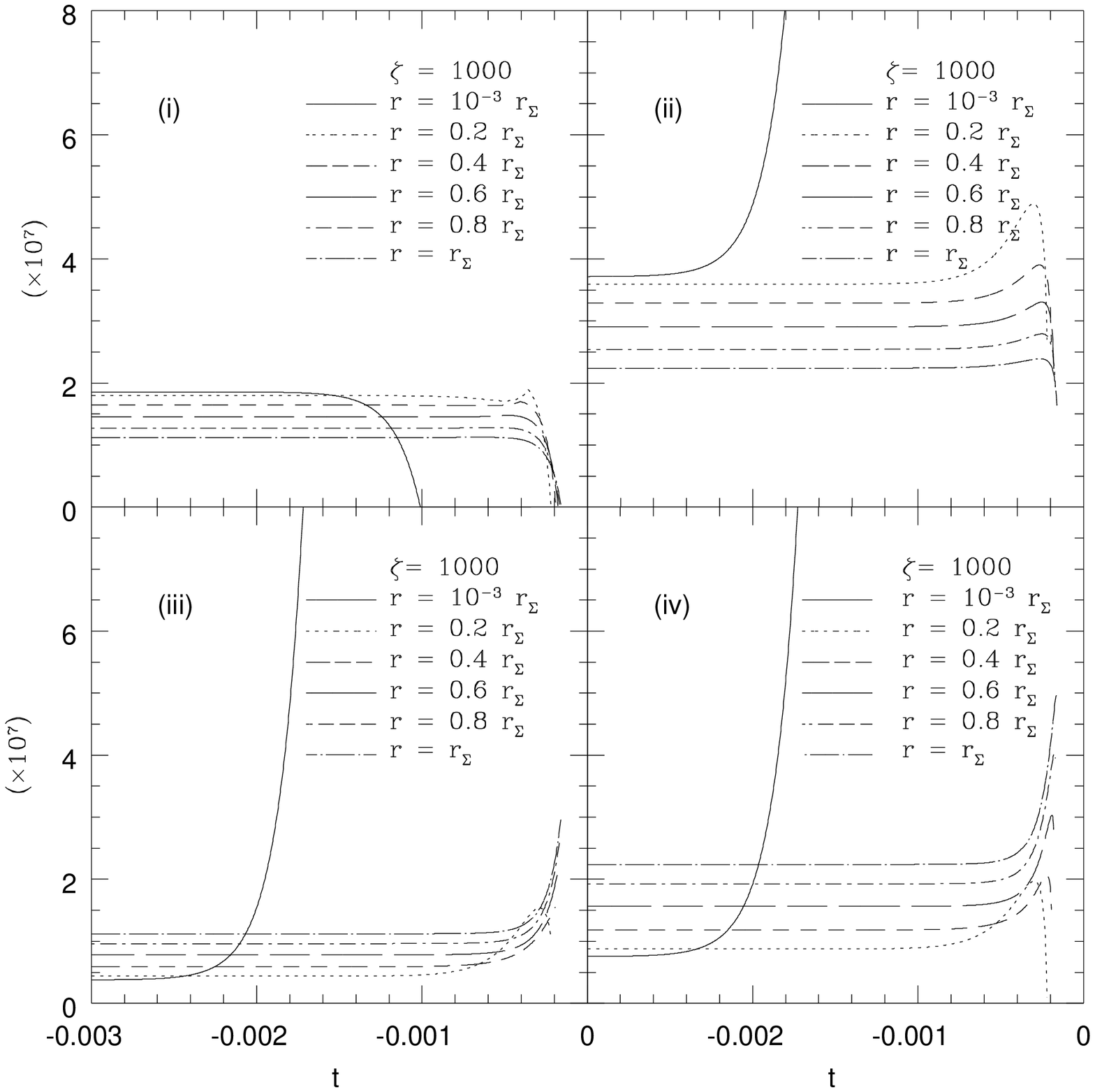}{The energy conditions (\ref{eq:gcr1})-(\ref{eq:dcr1}), 
for the model without bulk  viscosity, where $\zeta=1000$.  The time is in units 
of seconds and all the others quantities are in units of sec$^{-2}$.}{i-ivb}

\myfigure{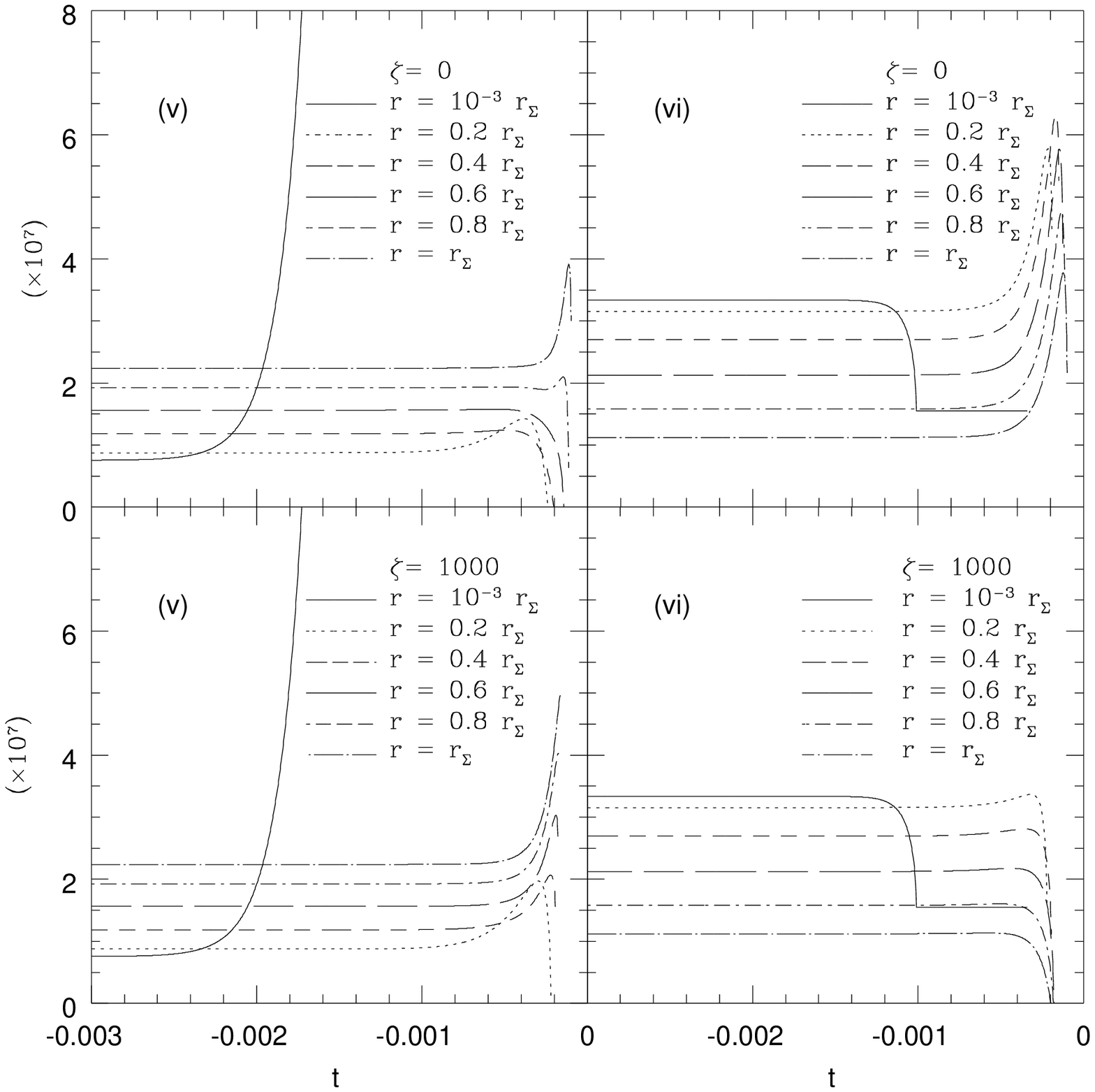}{The energy conditions (\ref{eq:dcr2})-(\ref{eq:scr1}), 
for the model with or without bulk  viscosity, where $\zeta=0$ and $\zeta=1000$.  The time is in units 
of seconds and all the others quantities are in units of sec$^{-2}$.}{v-viab}

In order to verify the energy conditions, we have plotted the time evolution
of all the conditions,
for several radii and for two values of $\zeta$ (0 and 1000),
as we can see in the figures \ref{i-iva}, \ref{i-ivb} and \ref{v-viab}. 
For the sake of comparison with the model $\zeta \ne 0$, we have plotted all
the conditions (\ref{eq:gcr1})-(\ref{eq:scr1}) for $\zeta = 0$.

From the figures \ref{i-iva}($i$) and \ref{i-ivb}($i$) we can conclude that only the 
inequality [$|\mu+p-\zeta \Theta|-2|\bar q| \ge 0$] is not satisfied during all
the collapse and for any radius.  This inequality is not satisfied for the 
innermost radii ($r \le 0.2r_{\Sigma}$) and for the latest stages of the 
collapse.  The condition (\ref{eq:scr1}) is not satisfied for $r < 0.2r_\Sigma$
[figure \ref{v-viab}($vi$)] because the inequality
(\ref{eq:gcr1}) [$\Delta \ge 0$] is not satisfied for these radii and for the 
latest stages of the collapse.  

\section{Physical Results}

As in previous papers
\cite{Chan97, Chan98a, Chan00, Chan01, Nogueira04}, 
we have calculated several physical quantities, as the total energy entrapped 
inside the $\Sigma$ surface, the total luminosity perceived by an observer 
at rest at infinity and the effective adiabatic index, and we have compared
them to the respective non-viscous ones.

From equation (\ref{eq:ms}) we can write using (\ref{eq:art})-(\ref{eq:crt}) and
(\ref{eq:a0})-(\ref{eq:p0a}) that

\begin{eqnarray}
{m \over m_0} &=& {f \over {16r_{\Sigma}^6(1-r_{\Sigma}^2)h^2}} \left[m_0^2 (1+r_{\Sigma}^2)^8 \dot f^2 h^2 + \right. \nonumber \\
& & \left.+ 4r_{\Sigma}^4 (1-r_{\Sigma}^4)^2(h^2-f^2) + 16r_{\Sigma}^6(1-r_{\Sigma}^2)^2f^2 \right],
\label{eq:mf}
\end{eqnarray}
where

\begin{equation}
m_0=-\left[ {r^2B'_0} + {r^3B'^2_0 \over {2B_0}} \right]_{\Sigma}.
\label{eq:m0}
\end{equation}

We can see from figure \ref{mass} that the total energy entrapped inside
the hypersurface $\Sigma$ vanishes at the time $-2.0\times10^{-5}$ s,
approximately.  This means that the star radiates all its mass during the
collapse and this explains why the apparent horizon never forms.
We can also observe from figure \ref{mass} that the mass inside $\Sigma$ is
equal for both models, with or without bulk viscosity.  This means that they
radiate the same amount of mass during the evolution.
In the figure \ref{massa} we can see the evolution of the mass from a
previous model \cite{Chan01}.  In contrast of the result of this work, the
former model radiates about 33\% of the total mass of the star, before the
formation of the black hole.

\myfigure{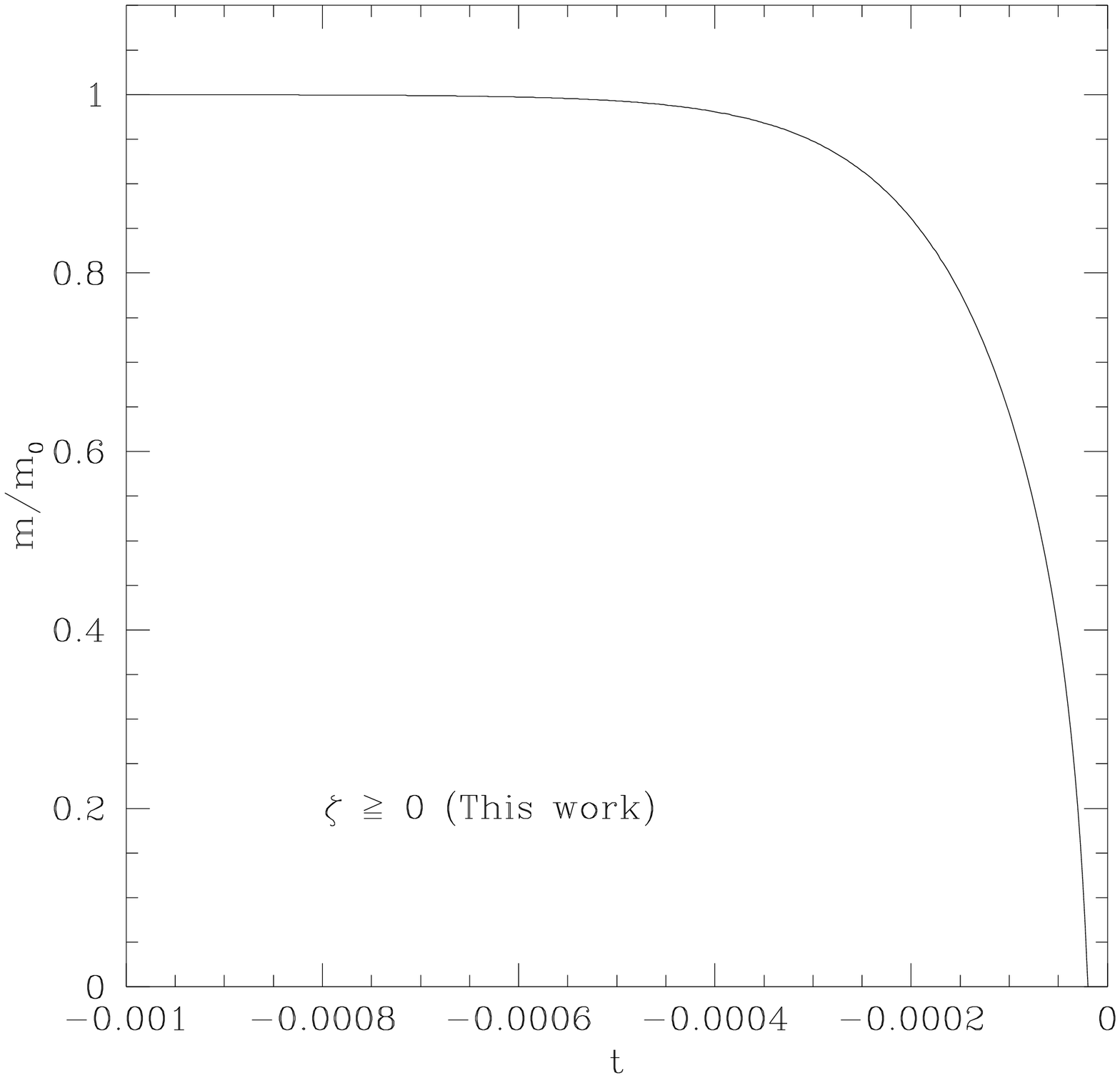}{Time behavior of the total energy entrapped inside
the surface $\Sigma$ for the models with or without bulk viscosity.  The 
time, $m$ and $m_0$ are in units of seconds. The symbols $\zeta \ge 0$ mean that
the plotted quantity is independent of $\zeta$.}{mass}

\myfigure{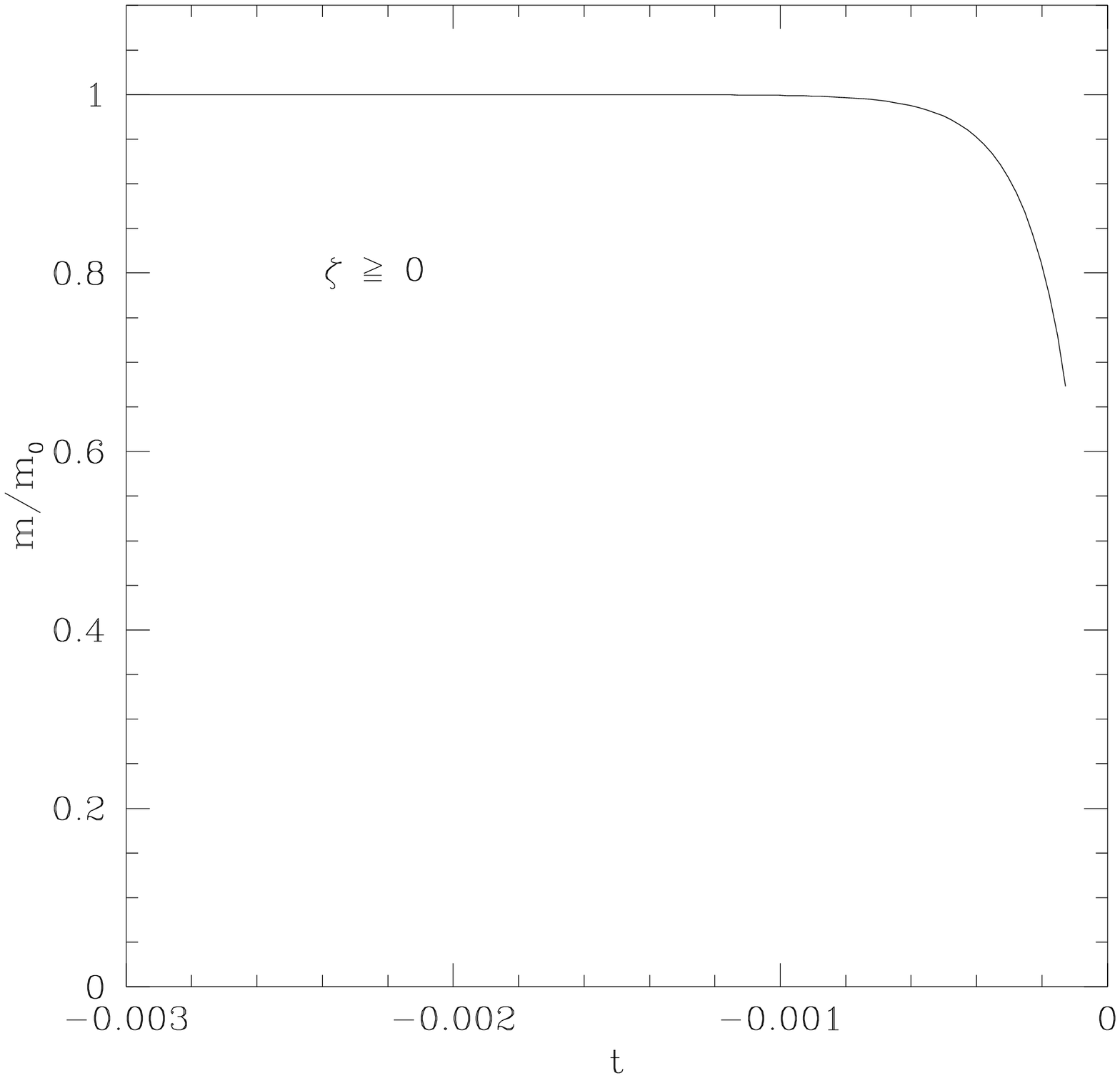}{Time behavior of the total energy entrapped inside
the surface $\Sigma$ for the models with or without bulk viscosity, 
from a previous model \cite{Chan01}.  The 
time, $m$ and $m_0$ are in units of seconds. The symbols $\zeta \ge 0$ mean that
the plotted quantity is independent of $\zeta$.}{massa}

Using the equations (\ref{eq:lsp}) and (\ref{eq:art})-(\ref{eq:crt}) we can
write the luminosity of the star as

\begin{eqnarray}
L_{\infty}&=&\kappa {{m_0^2(1+r_{\Sigma}^2)^4f^2} \over {8r_{\Sigma}^4}} \times \nonumber \\
& & \times \left[ \left({{1-r_{\Sigma}^2} \over {1+r_{\Sigma}^2}} \right)
\left( {f \over h} \right) + {{m_0(1+r_{\Sigma}^2)^3 \dot f} \over {2r_{\Sigma}^2(1-r_{\Sigma}^2)}} \right]^2 
\left[ p_{\Sigma} - \zeta\left( \frac{1+r^2_{\Sigma}}{1-r^2_{\Sigma}} \right)\left( 2\frac{\dot f}{f} + \frac{\dot h}{h} \right) \right].
\label{eq:lf}
\end{eqnarray}

We can see from figure \ref{lumin} that the luminosity perceived by an
observer at rest at infinity increases exponentially until the time
$-2.0\times10^{-5}$, when the total mass of the star vanishes.
In the figure \ref{lumina} we can see the evolution of the luminosity from a
previous model \cite{Chan01}.  In contrast of the result of this work, the
luminosity of the star also increases exponentially, reaching a maximum and after
it decreases until the formation of the black hole.

\myfigure{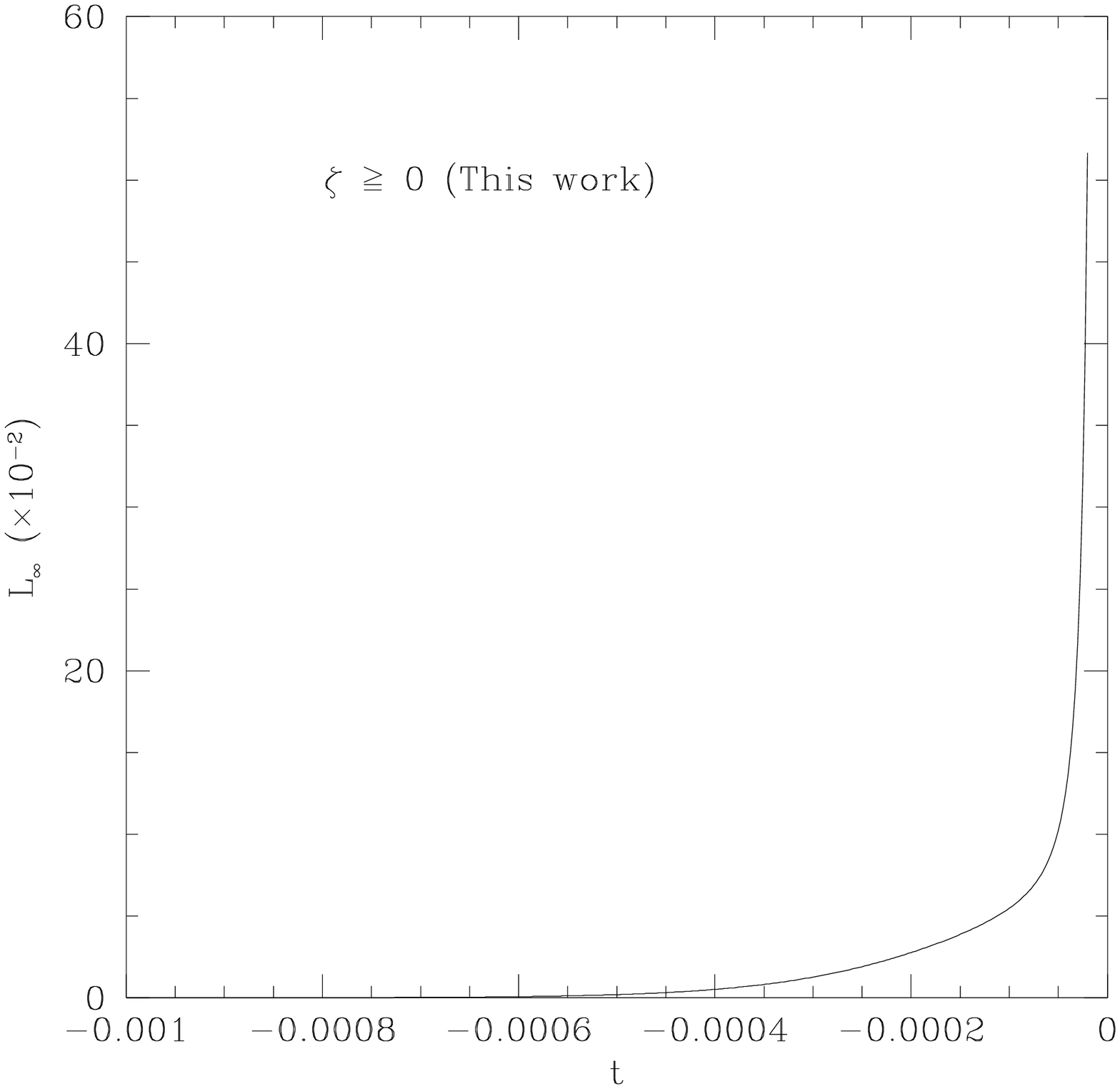}{Time behavior of the luminosity perceived by an
observer at rest at infinity for the models with or without bulk viscosity.  
The time is in units of second and the luminosity is dimensionless. 
The symbols $\zeta \ge 0$ mean that
the plotted quantity is independent of $\zeta$.}{lumin}

\myfigure{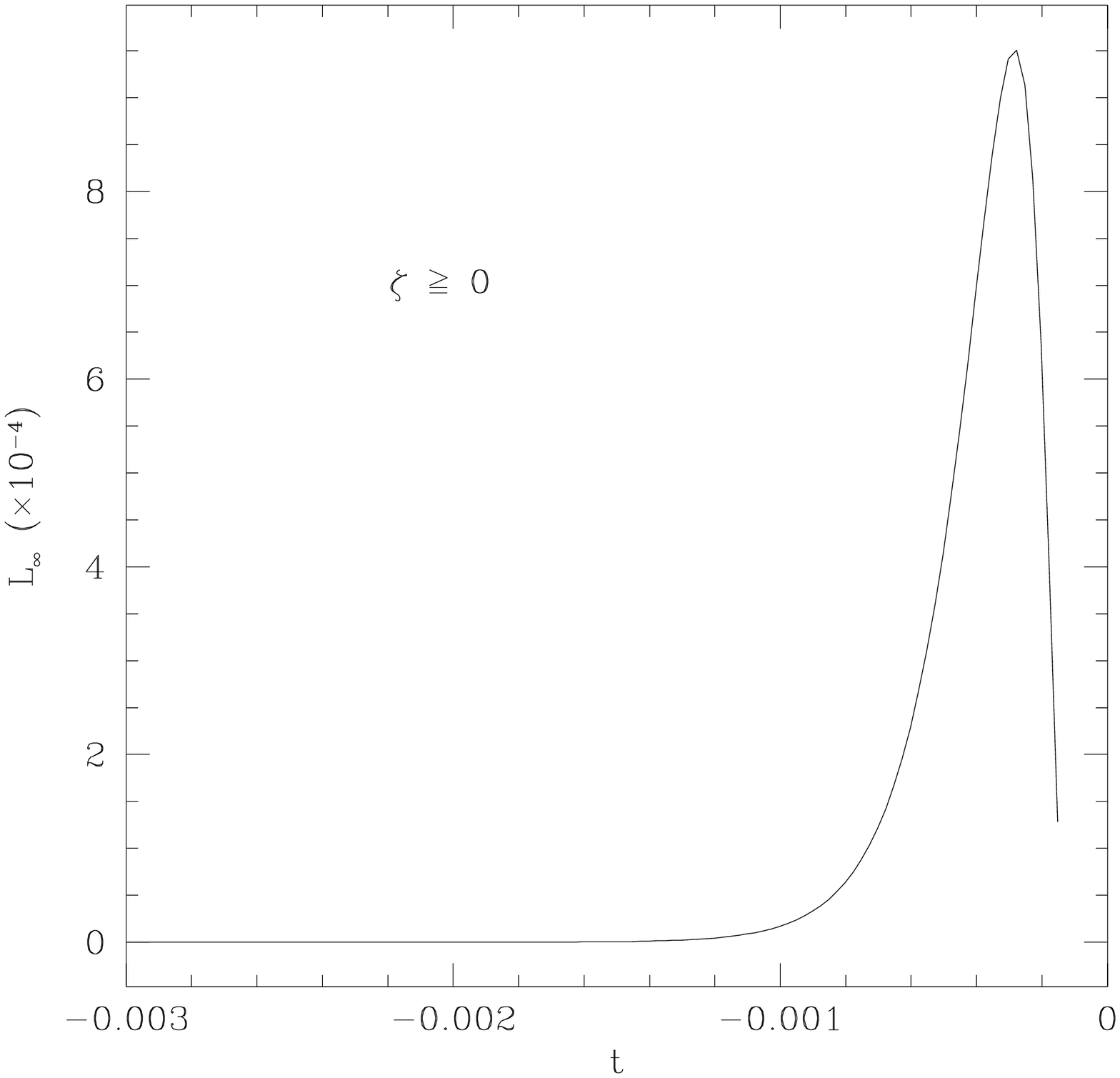}{Time behavior of the luminosity perceived by an
observer at rest at infinity for the models with or without bulk viscosity,
from a previous model \cite{Chan01}.  
The time is in units of second and the luminosity is dimensionless. 
The symbols $\zeta \ge 0$ mean that
the plotted quantity is independent of $\zeta$.}{lumina}

The effective adiabatic index can be calculated using the equations
(\ref{eq:mu})-(\ref{eq:p}), (\ref{eq:ft}) and 
(\ref{eq:a0})-(\ref{eq:p0a}). Thus, we can write that 

$$\Gamma_{\rm eff}=
\left[ {\partial (\ln p)} \over {\partial (\ln \mu)} \right]_{r=const} 
= \left( {\dot p} \over {\dot \mu} \right) \left( \mu \over p \right) = $$ 

\[ =\left\{ \left\{ \left[ 288r^2e(r)+12d(r)k(r) \right]f^2+6ac(r)k(r)hf\dot f -6 \kappa \zeta A_0 c(r)k(r)(h^3+h)f^2/a_0 \right\} f\dot h + \right. \]
\[ + c(r)k(r)j(r,t)f\left[3h^2\dot f^2+bh^2(1-f^2)\right] + \]
\[ \left. + k(r)h\dot f\left[c(r)(12bh^2+aj(r,t)f^2)-12d(r)h^2\right] \right\} \times \]
\[ \times 2^{-1}\left\{ \left\{24r^2d(r)f^2+k(r)\left[c(r)(3h^2\dot f^2+afh\dot f+bh^2(1-f^2))- \right. \right. \right. \]
\[ - \left. \left. 2d(r)f^2\right] \right\} f\dot h + c(r)k(r)(3h^3\dot f^3+ah^2f\dot f^2) + \]
\[ \left. + k(r)\dot f\left[bc(r)h^3(1-f^2)+2d(r)h^3\right]-2c(r)k(r)f^2\dot f h \dot h(1+h^2)/a_0 \right\}^{-1} \times \]
\[ \times \left\{ 12r^2d(r)f^2+k(r)\left[c(r)h^2\dot f^2+d(r)(h^2-f^2)\right] + \right. \]
\[ + \left. 2c(r)k(r)f\dot f h\dot h \right\} \times \]
\[ \times \left\{ 72r^2e(r)f^2+k(r)\left\{c(r)j(r,t)hf\dot f+3[c(r)b-d(r)] \times \right. \right. \]
\begin{equation}
\times \left. \left. (h^2-f^2)\right\} - c(r)k(r)l(r,t)f^2\dot h \right\}^{-1},
\label{eq:gamaf}
\end{equation}
where
\begin{equation}
c(r)=r^2m^2_0(1+r^2_{\Sigma})^8,
\label{eq:gamaf1}
\end{equation}
\begin{equation}
d(r)=r^6_{\Sigma}g^2(r),
\label{eq:gamaf2}
\end{equation}
\begin{equation}
e(r)=r^6_{\Sigma}(r^2_{\Sigma}-r^2)g(r),
\label{eq:gamaf3}
\end{equation}
\begin{equation}
k(r)=(1+r^2)^2,
\label{eq:gamaf4}
\end{equation}
\begin{equation}
j(r,t)=3 a + 6 \kappa \zeta A_0h.
\label{eq:gamaf5}
\end{equation}
and
\begin{equation}
l(r,t)=3 a_ob - 3 \kappa \zeta A_0h.
\label{eq:gamaf6}
\end{equation}

\myfigure{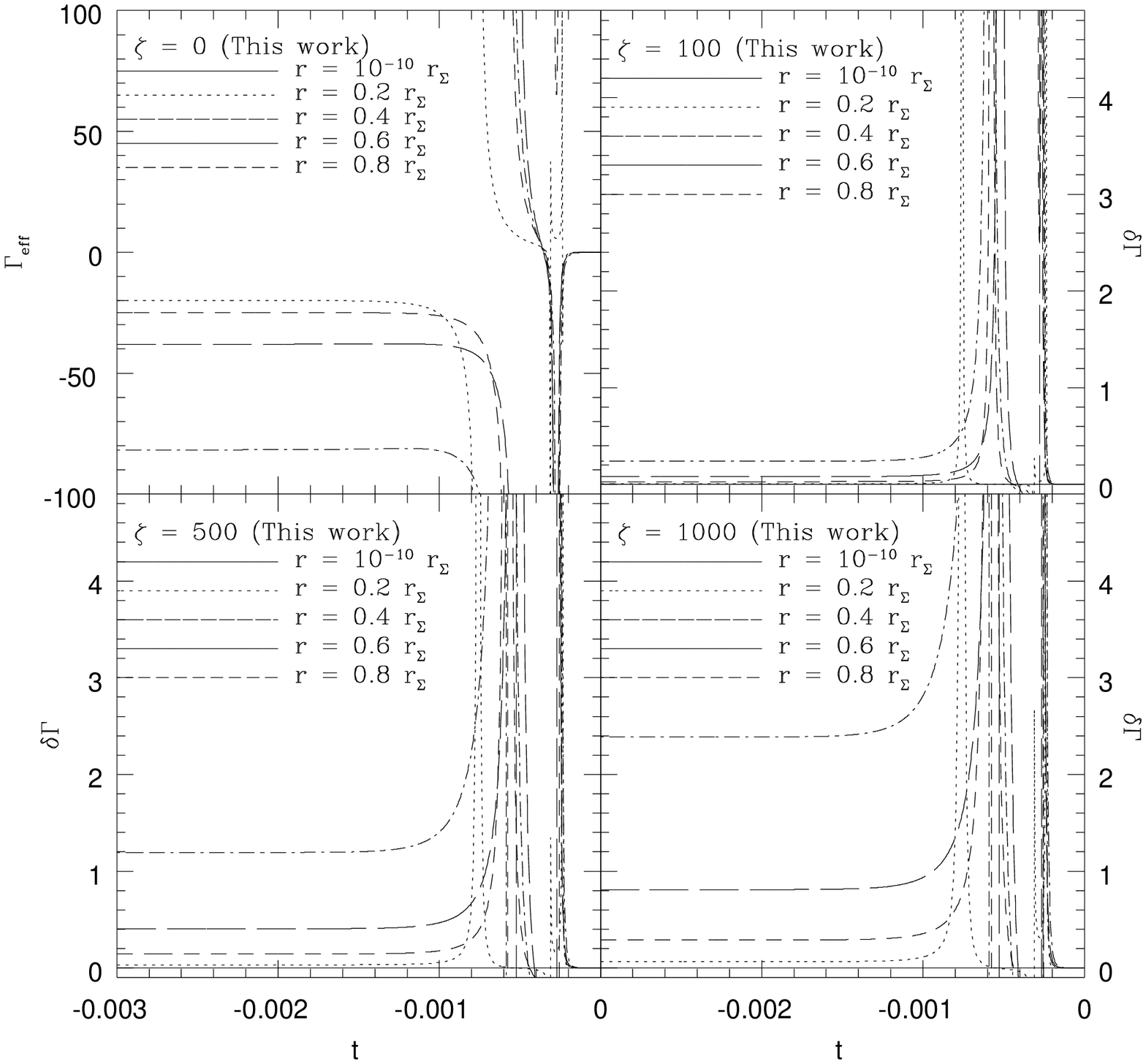}{Time behavior of the effective adiabatic index
$\Gamma_{\rm eff}$ for four values of $\zeta$.  The quantity $\delta \Gamma$ is 
defined as $\Gamma_{\rm eff}(\zeta = 0) - \Gamma_{\rm eff}(\zeta \ne 0)$.
The time is in units of seconds, $\Gamma_{\rm eff}$ and
$\delta \Gamma$ are dimensionless.}{gama}

\myfigure{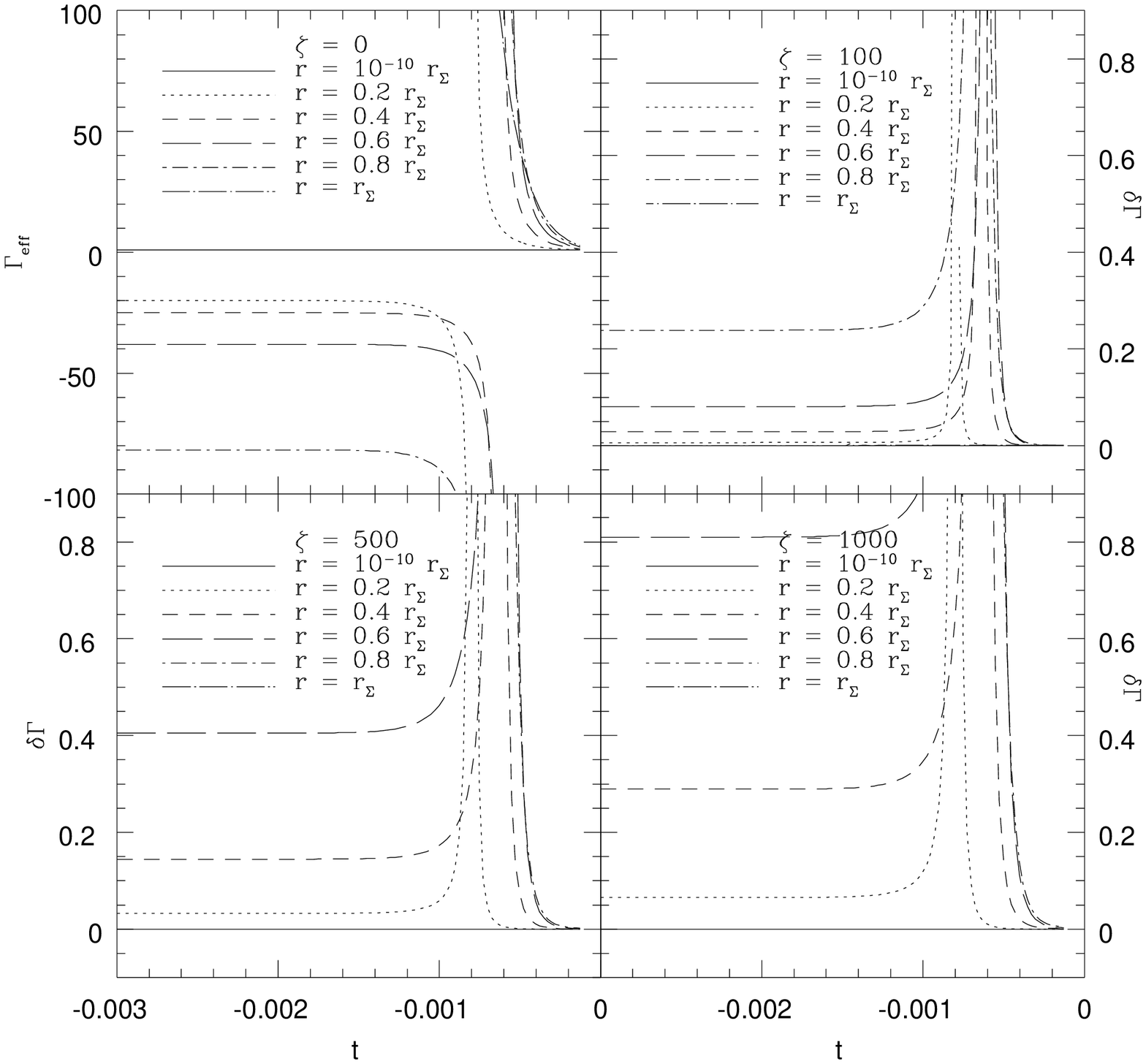}{Time behavior of the effective adiabatic index
$\Gamma_{\rm eff}$ for four values of $\zeta$, from a previous model \cite{Chan01}.  
The quantity $\delta \Gamma$ is 
defined as $\Gamma_{\rm eff}(\zeta = 0) - \Gamma_{\rm eff}(\zeta \ne 0)$.
The time is in units of seconds, $\Gamma_{\rm eff}$ and
$\delta \Gamma$ are dimensionless.}{gama1}

Comparing the figures for $\Gamma_{\rm eff}$ ($\zeta = 0$ and $\zeta \ne 0$) we can see that the time  
evolution of the effective adiabatic indices are not very different graphically.
This is the reason to plot the quantity 
$\delta \Gamma=\Gamma_{\rm eff}(\zeta = 0)-\Gamma_{\rm eff}(\zeta \ne 0)$ 
instead of $\Gamma_{\rm eff}$ for the $\zeta \ne 0$ models.
We can note in figure \ref{gama} ($\zeta=0$) that shortly before the 
peak of luminosity (see figure \ref{lumin}) there is a large discontinuity
in $\Gamma_{\rm eff}$ due mainly to the behavior of the pressure.
The effect of the viscosity is to increase much more these discontinuities.

In the figure \ref{gama1} we can see the evolution of the effective adiabatic indices from a
previous model \cite{Chan01}. We can note comparing it with the figure \ref{gama} that 
the effective adiabatic index
diminishes due to the bulk viscosity, thus increasing the instability of
the system, in both models, in the former paper \cite{Chan01} and in this work.
This characteristic might be model independent.

Finally, models of radiating viscous spheres have been presented by Herrera,
Jim\'enez and Barreto \cite{Herrera89}. This
work is particularly relevant for the proposed discussion
because the conclusion concerning the effective adiabatic index is the same
in both cases. Namely, an increasing of the critical adiabatic index
required for stability \cite{Chan94}, or equivalently, a decreasing of the 
effective
adiabatic index, induced by viscosity. Since the models considered in each
case are completely different, we suggest that
this effect seems to be model independent.

\section{Conclusions}

A new model is proposed to a collapsing star consisting of an
anisotropic fluid with bulk viscosity, radial heat flow and outgoing
radiation.  In a previous paper \cite{Chan03} one of us has 
introduced a function time dependent
into the $g_{rr}$, besides the time dependent metric functions
$g_{\theta\theta}$ and $g_{\phi\phi}$.  We have generalized
this previous model by introducing bulk viscosity and we have compared it to the
non-viscous collapse.

The behavior of the density, pressure, mass, luminosity and the
effective adiabatic index was analyzed. We have also compared to the case
of a collapsing fluid with bulk viscosity of another previous model \cite{Chan01}, 
for a star with 6 $M_{\odot}$.

As we have shown the black hole is never formed
because the apparent horizon formation condition is never satisfied.
This could be interpreted as the formation of a naked singularity, as
Joshi, Dadhich and Maartens \cite{Joshi02} have suggested.  However
this is not the case because the star radiates all its mass before it
reaches the singularity at $r=0$ and $t=0$.  Not even a marginally naked
singularity is formed by the same reason, since in this case the apparent
horizon should coincide with the singularity at $r=0$ and $t=0$.

The density and pressure have negative values although physically this
could be considered unreasonable.  However, due to the heat flow (the term
$\Delta$) the energy conditions are partially satisfied.

The pressure of the star, at the beginning of the collapse, is isotropic but
due to the presence of the bulk viscosity the pressure becomes more and more
anisotropic.  

The star radiates all its mass during the collapse and this explains 
why the apparent horizon never forms.
In contrast of the result of this work, the
former model radiates about 33\% of the total mass of the star, before the
formation of the black hole.

An observer at infinity will see a radial point source radiating
exponentially until reaches the time of maximum luminosity and
suddenly the star turns off because there is no more mass in order to be
radiated.
In contrast of the former model \cite{Chan01}
where the luminosity also increases exponentially, reaching a maximum and after
it decreases until the formation of the black hole.

The effective adiabatic index has a very unusual behavior because we have
a non-adiabatic regime in the fluid due to the heat flow.  The index becomes
negative since the hydrodynamic pressure and the density may become negative.
Besides, in this case, neither the density is the measure of the
total energy density of a given piece of matter nor the hydrodynamic
pressure the only opposing contraction \cite{Barreto92}.
The effective adiabatic index
diminishes due to the bulk viscosity, thus increasing the instability of
the system, in both models, in the former paper \cite{Chan01} and in this work,
showing that this characteristic might be model independent.

\bigskip
\bigskip
\noindent {\bf ACKNOWLEDGMENTS}
\bigskip

The author (RC) acknowledges the financial
support from FAPERJ (no. E-26/171.754/2000, E-26/171.533/2002 and
E-26/170.951/2006) and from
Conselho Nacional de Desenvolvimento Cient\'{\i}fico e Tecnol\'ogico - CNPq -
Brazil.
The author (GP) also acknowledges the financial support from CAPES.

\end{document}